\documentclass[aps,prb,twocolumn,showpacs,amsmath,superscriptaddress]{revtex4-1}

\usepackage{amsmath}
\usepackage{amssymb}
\usepackage{graphicx}
\usepackage{bbm}
\usepackage{bbold}
\usepackage{color}

\begin{document}

\title{{Tunable Berry curvature,  valley and spin Hall effect in Bilayer MoS$_2$}}

\author{Andor Korm\'anyos}
\thanks{e-mail: andor.kormanyos@uni-konstanz.de}
\affiliation{Department of Physics, University of Konstanz, D-78464 Konstanz, Germany}

\author{Viktor Z\'olyomi}
\affiliation{School of Physics and Astronomy, University of Manchester, Manchester M13 9PL, United Kingdom}

\author{ Vladimir I. Fal'ko}
\affiliation{School of Physics and Astronomy, University of Manchester, Manchester M13 9PL, United Kingdom}

\author{Guido Burkard}
\affiliation{Department of Physics, University of Konstanz, D-78464 Konstanz, Germany}

\begin{abstract}
The chirality of electronic Bloch bands is responsible for many intriguing properties of 
 layered two-dimensional materials.
 We show that in bilayers of transition metal dichalcogenides (TMDCs), unlike in few-layer graphene and 
 monolayer TMDCs, both intra-layer and inter-layer couplings give important 
 contributions to the Berry curvature in  the $K$ and $-K$ valleys of the Brillouin zone. 
 The inter-layer contribution leads to the stacking dependence of the Berry curvature and 
 we point out the differences between the commonly available 3R type and 2H type bilayers.
 Due to the inter-layer contribution the Berry curvature becomes highly tunable in double gated devices. 
 We study the dependence of the valley Hall  and spin Hall effects on the stacking type and external 
 electric field. Although the valley and spin Hall conductivities are not quantized, in MoS$_2$ 2H bilayers  
 they may change sign as a function of the external electric field which is reminiscent of the behaviour of 
 lattice Chern insulators. 
\end{abstract}


\maketitle

\section{Introduction}
\label{sec:intro}

The valley degree of freedom has  recently attracted a large interest in monolayers of 
group-VI transition metal dichalcogenides (TMDCs). 
This is in good part due to the fact that monolayer TMDCs exhibit circular optical 
dichroism, that is,  the valleys at the $\pm K$ point of the Brillouin zone (BZ)  can be 
directly addressed by left or right circularly polarized light\cite{Cui,Heinz,Cao,Urbaszek}. 
A related phenomenon,  called the valley-Hall effect, has also been 
{demonstrated\cite{monolayer-valleyH} in monolayer MoS$_2$, which can be traced to}
{the chirality of the electronic  Bloch bands\cite{DiXiao-valley,WangYao-valley,WangYao-TMDCmonolayer}.}
The Berry curvature\cite{Berry-phase-review} properties of bilayer TMDCs have received very limited 
attention so far\cite{symmetry-tuning},
{in part, due to the uncertainty about the position of the band edges in the Brillouin zone 
that one can find in the existing literature\cite{monolayer-valleyH,DiXiao-valley,WangYao-valley,WangYao-TMDCmonolayer}.}
A better understanding of the Berry curvature properties would be important 
in light of  recent reports\cite{valley-Hall-BLMoS2,Morpurgo-3R-bilayer} on the valley-Hall effect in bilayer 
{MoS$_2$, and the purpose of this work is to analyse the topological properties of bilayer 
TMDCs (BTMDCs).}

Because of the recent experimental progress\cite{symmetry-tuning,Iwasa,Arita, 
valley-Hall-BLMoS2,XiangSheng,Morpurgo-3R-bilayer}, we will concentrate  on bilayer MoS$_2$ in the following, 
but many of our findings are equally valid for other {BTMDCs} such as MoSe$_2$, WS$_2$, and WSe$_2$. 
{The focus of the present study is on the competition between the contributions 
towards Berry curvature of electron bands in BTMDCs coming from the intrinsic properties of the monolayers 
and a part generated by the inter-layer coupling.} Thus, BTMDCs are markedly different 
from gapped bilayer graphene or monolayer TMDCs, where only one of the 
contributions is finite\cite{WangYao-TMDCmonolayer,Morpurgo-BLG}. Because of the inter-layer contribution, the 
Berry curvature is tunable by moderately strong external electric fields. 
Moreover,  we show that the stacking of the  monolayer constituents in BTMDCs affects 
the  Berry curvature  and different stackings have  Berry curvature properties. 
These topological differences can already be understood if  spin-orbit coupling (SOC) is neglected.  
Nevertheless, we will also analyse  the effect of SOC  on the band structure, on the Berry curvature  
and on certain transport properties.  
The finite Berry curvature leads to valley and spin Hall conductivities which 
depend on the stacking and on the presence/absence of inversion symmetry in the system. 
{As we will show below, the interplay of intrinsic SOC, the layer degree of freedom and an 
external electric field can lead to an interesting effect: the valley and spin Hall 
conductivities  change sign as a function of the external electric field. }

Generally, the presence/absence  of inter/intra-layer Berry curvature contributions 
and the effect of different stacking is a relevant question for  all layered materials, including, e.g., 
heterostructures of different monolayer TMDCs obtained by layer-by-layer growth\cite{Ajayan} or 
artificial alignment\cite{TobiasKorn}. BTMDCs, in addition,  present a novel, rich playground for 
valley and spin related phenomena. 

\begin{figure*}[htb]
  \includegraphics[scale=0.6]{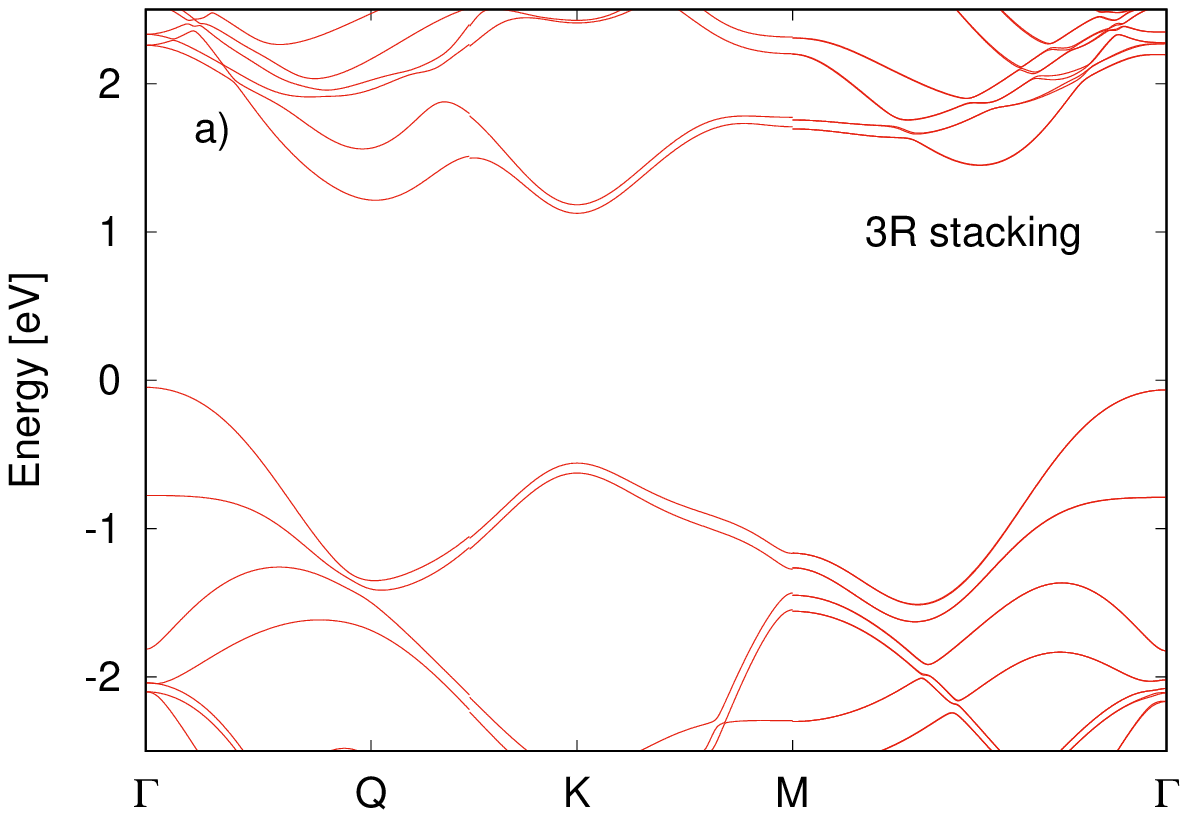}
  \includegraphics[scale=0.6]{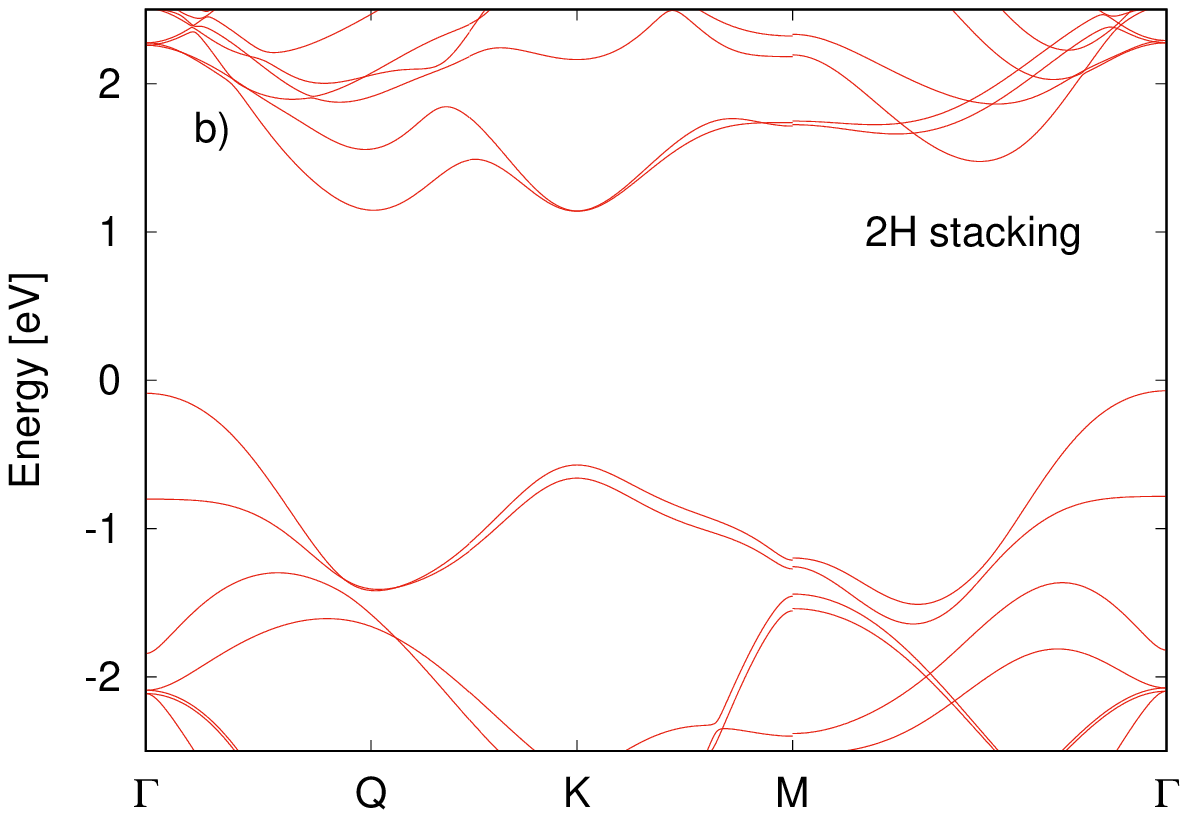}
 \includegraphics[scale=0.35]{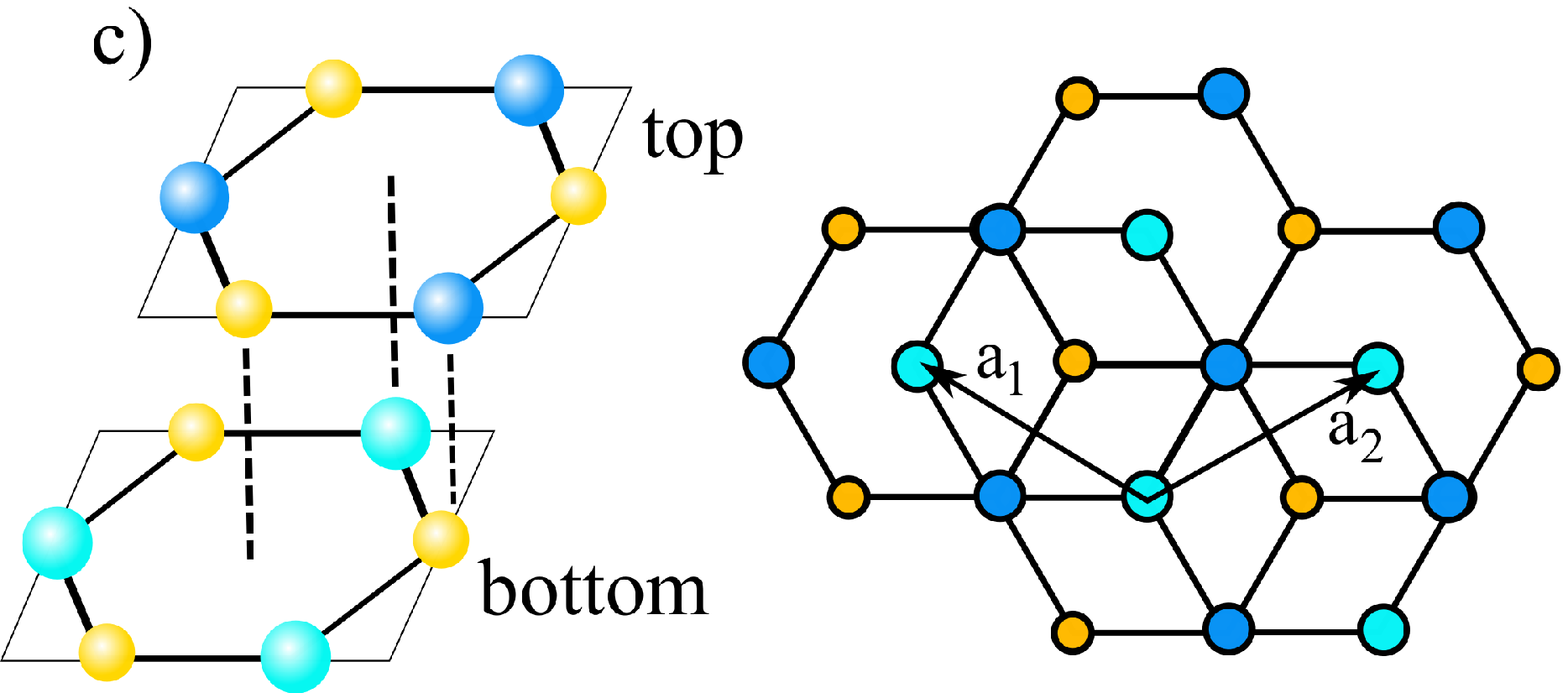}\hspace{1.5cm}
 \includegraphics[scale=0.35]{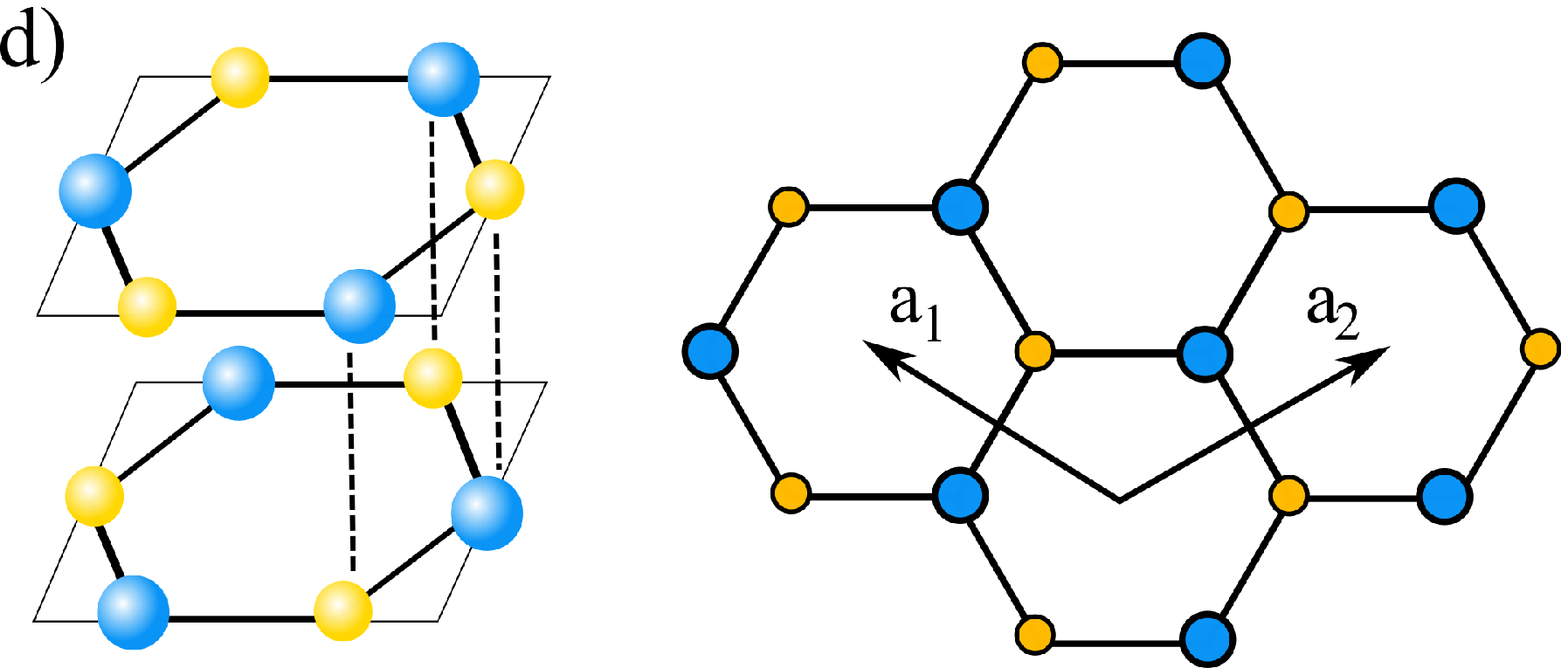}
 \caption{{a) Band structure showing the CB and the VB of 3R stacked bilayer MoS$_2$ along the $\Gamma-K-M-\Gamma$ direction 
             of the BZ from DFT calculations. b) The same for 2H stacked bilayer MoS$_2$. 
             The spin-orbit coupling is neglected in a) and b).
           c) Schematic crystal structure of 3R stacked bilayer TMDCs  in side view and in top view. 
           d) Schematic crystal structure of 2H stacked bilayer TMDCs in side view, and in top view.
           In c) and d) the monolayers are shown as simple hexagonal lattices with two inequivalent sites, while
           $\mathbf{a}_1$ and $\mathbf{a}_2$ denote the lattice  vectors.}            
 \label{fig:cryst-struct-bilayer}}
\end{figure*}


\section{$\mathbf{k}\cdot\mathbf{p}$ Hamiltonian in the $\pm K$ valleys}
\label{sec:kp-Ham}

{There are two naturally occurring stable phases of bulk TMDCs with an underlying hexagonal symmetry of their
lattice structure\cite{Wilson}. The most common one is the so-called 2H polytype, where the unit cell contains two 
monolayer units and the bulk is inversion symmetric. Some layered TMDCs, among others MoS$_2$, can also exist 
in the 3R polytype, where the unit cell contains three monolayers and inversion symmetry is broken in the bulk. 
Bilayer samples can be exfoliated from both bulk phases and we will refer to them as 2H and 3R stacked bilayers. }

We start our discussion by introducing the $\mathbf{k}\cdot\mathbf{p}$ Hamiltonians for 3R and 2H stacked 
bilayers. We will focus on the $\pm K$ valleys in the BZ  because 
in our DFT calculations [see Figures \ref{fig:cryst-struct-bilayer}(a)-(b)] 
the band edge  in the conduction band can be found at these point, therefore they are  experimentally relevant. 
We will briefly discuss the $Q$ valleys in Section~\ref{sec:Q-pt-main}. The
main differences between the Berry curvature properties of 3R and 2H bilayer TMDCs are orbital effects and 
therefore we neglect the SOC in the present Section.
The discussion of the important effects of the SOC are deferred to Sections \ref{sec:SOC} and \ref{sec:valley-Hall}.

\subsection{3R bilayers}
\label{subsec:3R-K-kp}

We start our discussion with 3R stacked bilayer MoS$_2$, which can be 
exfoliated from the 3R bulk polytype\cite{Iwasa,Arita,XiangSheng,Morpurgo-3R-bilayer}. 
3R bilayers are non-centrosymmetric (the symmetry of the crystal structure is described 
by the point group $C_{3v}$) and therefore, in contrast to 2H stacked bilayers (see below)  
one can expect interesting Berry curvature properties even if no external electric field is applied. 
We first introduce a $\mathbf{k}\cdot\mathbf{p}$ model for this system which differs in  important 
details from the one used recently in Ref.~\onlinecite{Xiaoou} (see Appendix \ref{subsec:K-pt-3R} for
the derivation of this model). 
The electronic properties at the $\pm K$ point of the  BZ can be  succinctly captured 
by the following simplified  $\mathbf{k}\cdot\mathbf{p}$ Hamiltonian: 
\begin{equation}
H^{3R}_{K} =
\left(
\begin{array}{cccc}
 \varepsilon_{cb}^{b} & \gamma_3\,q_{+} &  \gamma_{cc}\,q_{-} &  0\\
 \gamma_3\, q_{-} & \varepsilon_{vb}^{b} &  0    &  \gamma_{vv}\,q_{-}\\
 \gamma_{cc}\,q_{+} &  0  & \varepsilon_{cb}^{t}  & \gamma_{3} q_{+} \\
 0   &  \gamma_{vv} q_{+}  &    \gamma_{3} q_{-}  &  \varepsilon_{vb}^{t}      
\end{array}
\right)
\label{eq:H-3R-4dim}
\end{equation}
Here $q_{\pm}=\tau q_x \pm i q_y$ denotes the wavenumber measured from the $K$ (or $-K$)
point of the  BZ and $\tau=\pm 1$ is the valley index. Higher order terms in $q_{\pm}$, which 
appear in the $\mathbf{k}\cdot\mathbf{p}$ model of monolayer TMDCs\cite{PRB-paper, Asgari,2DMaterials-paper} 
have been neglected here. The band-edge energies of the CB and VB
in the bottom (top) layers are denoted by $\varepsilon_{cb}^{b}$ ($\varepsilon_{cb}^{t}$)
and $\varepsilon_{vb}^{b}$ ($\varepsilon_{vb}^{t}$). 
The layer index bottom (b) and top (t) are assigned to the bands based on the 
localization of the corresponding Bloch-wave function to one of the layers.  
We note that explicit density functional theory (DFT) wave function calculations for the bilayer case can be found 
in Ref.~\onlinecite{XiangSheng}, while for the bulk 3R polytype in Ref.~\onlinecite{Iwasa}. 
Our definition of the layer index is shown in Fig.~\ref{fig:cryst-struct-bilayer}(c): in the bottom monolayer the Mo 
atom does not have a S neighbour atom directly above it, while for the Mo atom in the top layer there is a S atom neighbour 
belonging to the bottom monolayer. Since in the monolayers the  atomic Mo $d$ orbitals have the largest 
weight in the conduction and the valence band (CB and VB) at the $\pm K$ points one may expect 
that this difference in the  atomic environment of the two Mo atoms can lead to different crystal 
field splittings in the two layers and hence it may  affect the band structure of the bilayers. 
This is indeed what we can deduce from our DFT band structure calculations, i.e., that  
$\varepsilon_{cb\,(vb)}^{(b)} > \varepsilon_{cb\,(vb)}^{(t)}$, see also Appendix \ref{subsec:K-pt-3R}. 
(We performed our DFT calculations using the VASP code\cite{VASP}, for further 
details see Ref.~\onlinecite{Viktor-DFT}).
Defining the band-edge energy differences 
$\delta E_{cc}^{}= (\varepsilon_{cb}^{b}-\varepsilon_{cb}^{t})/2$ and 
$\delta E_{vv}^{}= (\varepsilon_{vb}^{b}-\varepsilon_{vb}^{t})/2$, 
our DFT band structure calculations suggest that $\delta E_{cc}^{}\neq\delta E_{vv}^{}$,
meaning that there is a small difference of about $10$\,meV between the band gaps 
of the bottom and 
 the top layer. This energy difference 
will be neglected  in the following  as this does not affect any of the 
main conclusions of the Berry curvature calculations.
We denote therefore by  $\delta E_{cc}^{}=\delta E_{vv}^{}:=\delta E_{ll}^{}$ the inter-layer band-edge energy
difference and use the notation $\delta E_{bg}^{}=E_{bg}^{}/2$ 
for half of the monolayer bandgap. 
We use $\gamma_3$ for the intra-layer coupling of the CB and VB,  and $\gamma_{cc}$ ($\gamma_{vv}$) is 
the inter-layer couplings between the CBs (VBs) of the two monolayers. 
{The numerical value of $\gamma_3$ can, in principle, be somewhat different in the two layers, but 
we neglect this effect and use the monolayer value.} 
The coupling constants $\gamma_{cc}$ and $\gamma_{vv}$ can be estimated by fitting the eigenenergies of $H^{3R}_{K}$ to 
the DFT band structure (see Table \ref{tbl:params}). One can see from Eq.~(\ref{eq:H-3R-4dim}) that for 
$\mathbf{q}=0$ the two layers are decoupled, in agreement with previous results\cite{Arita} for the 3R bulk form.


\subsection{2H bilayers}
\label{subsec:2H-K-kp}
We now compare the results in Section \ref{subsec:3R-K-kp} with the corresponding ones for 
2H stacked bilayer MoS$_2$ which derives from the 2H polytype [see Fig.~\ref{fig:cryst-struct-bilayer}(b)]. 
The $\mathbf{k}\cdot\mathbf{p}$ Hamiltonian reads 
\begin{equation}
H^{2H}_{K} =
\left(
\begin{array}{cccc}
 \varepsilon_{cb}^{}+U_g & \gamma_3\,q_{+} &  \gamma_{cc}\,q_{-} &  0\\
 \gamma_3\, q_{-} & \varepsilon_{vb}^{}+U_g &  0    &  t_{\perp}\\
 \gamma_{cc}\,q_{+} &  0  & \varepsilon_{cb}^{}-U_g  & \gamma_{3} q_{-} \\
 0   &  t_{\perp}  &    \gamma_{3} q_{+}  &  \varepsilon_{vb}^{}-U_g      
\end{array}
\right)
\label{eq:H-2H-4dim}
\end{equation}
where $t_{\perp}$ is a momentum independent tunnelling amplitude between the VBs of the 
two layers and we included the possibility of an inter-layer potential difference  given by $\pm U_g$, 
which can be induced by a substrate or an external electric field. 
A similar  model,   which neglected the coupling between the CBs, was introduced in  
Refs.~\onlinecite{WangYao-bilayer,symmetry-tuning} (see the Appendix \ref{subsec:K-pt-2H} for further details.)  
We will show, however, that the coupling between the CBs gives an  important contribution to the Berry curvature.  
For $U_g=0$ the system  is inversion symmetric (the crystal symmetries are described by 
point group $D_{3d}$). 
At the $\pm K$ points  the two CBs are degenerate, while the  VBs are split 
due to the tunnelling amplitude $t_{\perp}$ [see Fig.~\ref{fig:cryst-struct-bilayer}(b)]. 
Away from  the $\pm K$ points the CBs are also split, for small $\mathbf{q}$ wavenumbers 
this splitting is mainly due to the interlayer coupling term $\gamma_{cc} q_{\pm}$.


\section{Berry curvature of BTMDCs}
\label{sec:Berry-curv}

\subsection{Numerical results and analytical approach}

The Berry curvature of band $n$ in a 2D material is defined by 
$\Omega_z(\mathbf{k})=\nabla_{\mathbf{k}}\times i\langle u_{n,\mathbf{k}}|\nabla_{\mathbf{k}}u_{n,\mathbf{k}}\rangle$, 
where  $u_{n,\mathbf{k}}$ is the lattice-periodic part of the Bloch wave functions. 
In the envelope function approximation $u_{n,\mathbf{k}}$ can be calculated from
a $\mathbf{k}\cdot\mathbf{p}$ Hamiltonian valid around a certain $\mathbf{k}$-space point. 
{Using the $\mathbf{k}.\mathbf{p}$ models introduced in Sections \ref{subsec:3R-K-kp} and \ref{subsec:2H-K-kp}, 
in the $\pm K$ valleys the $u_{n,\mathbf{k}}$ functions are  4-spinors that 
can be obtained by e.g., numerically diagonalizing $H^{3R}_{K}$ and $H^{2H}_{K}$ of 
Eqs.~(\ref{eq:H-3R-4dim}) and (\ref{eq:H-2H-4dim}), respectively.
We used these eigenstates and the approach introduced by Ref.~\onlinecite{Fukui} to  calculate 
the Berry curvature. }
The $\Omega_{z}(\mathbf{k})$ obtained for 3R and 2H bilayers is 
shown in Figures \ref{fig:Berry-curv-calc}(a) and (b), respectively 
(the material parameters used in these calculations are given in Table \ref{tbl:params}).   
For comparison, we also show the Berry curvature that can be obtained from a gapped-graphene model\cite{WangYao-TMDCmonolayer} 
which approximately describes the band structure of  individual monolayers in the limiting case when all inter-layer coupling 
terms in Eqs.~(\ref{eq:H-3R-4dim}) and (\ref{eq:H-2H-4dim})  are neglected. It is clear that the Berry curvature 
of both types of bilayer is substantially different from the monolayer case suggesting that 
inter-layer coupling may have an important role. 

To show this explicitly, we now derive an approximation for $\Omega_{z}^{}(\mathbf{k})$ which can make 
analytical calculations easier. 
As it is well known,  one can  use the Schrieffer-Wolff transformation\cite{Winkler-book}  
$e^{-S} H e^{S}$ of a Hamiltonian $H$ to eliminate coupling  terms  between  subsystems  of $H$
in order to obtain an effective Hamiltonian $\tilde{H}$ in the desired subspace.  
Here $S=-S^{\dagger}$ is an anti-Hermitian operator.
Denoting the eigenfunctions of $\tilde{H}$ by $\tilde{\Psi}$, the eigenfunctions $\Psi$ of the original 
Hamiltonian $H$ can be obtained by a back-transformation $\Psi=e^{S}\tilde{\Psi}$. 
By writing $e^{S}=\mathbb{1}+S+\frac{1}{2!}S^2+\dots$ a systematic approximation of 
$\Psi$ can be obtained if $\tilde{\Psi}$ is known. By using this approximation for $\Psi$ 
in the expression of $\Omega_z$, one finds 
\begin{equation}
 \Omega_z(\mathbf{k}) = \nabla_{\mathbf{k}}\times i \left\{
 \langle \tilde{\Psi}| \nabla_{\mathbf{k}} \tilde{\Psi}\rangle+
 \frac{1}{2}\langle \tilde{\Psi}| [\nabla_{\mathbf{k}}S^{},S^{}] |\tilde{\Psi} \rangle + \dots
 \right\}
 \label{eq:Omegaz-perturb}
\end{equation}
where $[A,B]$ denotes the commutator of $A$ and $B$. 
Although $S$ is usually not known exactly, one can write $S=S^{(1)}+S^{(2)}+\dots$ and  
explicit expressions for $S^{(i)}$ can be found  in e.g., Ref.~\onlinecite{Winkler-book}. 
In this way Eq.~(\ref{eq:Omegaz-perturb}) can be used to obtain a perturbation series 
for $\Omega_z$. 
One may write 
$
 \Omega_z \approx \Omega_z^{(0)} +\Omega_z^{(1,1)} + \dots
$
where $\Omega_z^{(0)}= \nabla_{\mathbf{k}} \times i\langle \tilde{\Psi}| \nabla_{\mathbf{k}} \tilde{\Psi}\rangle$ 
and 
$\Omega_z^{(1,1)}= \nabla_{\mathbf{k}} \times \frac{i}{2} 
\langle \tilde{\Psi}| [\nabla_{\mathbf{k}}S^{(1)},S^{(1)}] | \tilde{\Psi}\rangle$.

\subsection{Berry curvature of 3R and 2H bilayers}

In the case of 3R  bilayers one may choose as a subspace e.g., one of the layers and treat the inter-layer coupling 
as perturbation. 
This  corresponds to neglecting the inter-layer coupling in the wave function but retaining it in $S^{(1)}$.  
Using Eq.~(\ref{eq:Omegaz-perturb}) we find that $\Omega_{z}^{(b)}$  ($\Omega_{z}^{(t)}$) 
for  the bottom (top) layer can be written as
$\Omega_z^{(b)}\approx  \Omega_z^{(0)}-\Omega_z^{(1,1)}$ 
($\Omega_z^{(t)}\approx  \Omega_z^{(0)}+\Omega_z^{(1,1)}$), where  
\begin{subequations}
\begin{equation}
 \Omega_z^{(0)}(\mathbf{q})=\pm \frac{\tau}{2} \left(\frac{\gamma_3}{\delta E_{bg}}\right)^2
 \frac{1}{\left(1+\left(\frac{\gamma_3 |\mathbf{q}| }{\delta E_{bg}}\right)^2\right)^{3/2}},
 \label{eq:omega0-aa}
 \end{equation}
 \begin{equation}
 \Omega_z^{(1,1)}(\mathbf{q})\approx\frac{\tau}{(2\delta E_{ll})^2}
 \left(
 \lambda_1 \pm
 \frac{\lambda_2}{\left(1+\left(\frac{\gamma_3 |\mathbf{q}| }{\delta E_{bg}}\right)^2\right)^{1/2}}, 
 \right) \label{eq:omega11a-aa}
  \end{equation}
 \label{eq:omegaz-3r}
 \end{subequations}
Here $|\mathbf{q}|$ is the magnitude of $\mathbf{q}$, 
$\lambda_1=\gamma_{cc}^2+\gamma_{vv}^2$, $\lambda_2=\gamma_{cc}^2-\gamma_{vv}^2$ and 
the $+$ ($-$) sign corresponds to the CB (VB).  
$\Omega_z^{(0)}$ in Eq.~(\ref{eq:omega0-aa}) is the well known result for a 
gapped-graphene two-band model\cite{Berry-phase-review,WangYao-valley}, 
while Eq.~(\ref{eq:omega11a-aa}) 
is a correction due to the inter-layer coupling. 
The  first correction to Eq.~(\ref{eq:omegaz-3r}) is  $\sim \mathbf{q}^2$ but we found 
that for the wavenumber range of interest it is quite small. 

In 2H bilayers,  if  both inversion and time reversal 
symmetries are simultaneously present,   $\Omega_z$ vanishes\cite{Berry-phase-review}.  
However, a finite inter-layer potential $\pm U_{g}$ breaks inversion symmetry, opens a gap in the 
CB at the $\pm K$ point,  and causes $\Omega_z(\mathbf{q})$ to be non-zero. 
For the physically relevant case of $U_{g}\ll \delta E_{bg}$  
it proves to be useful to treat the intra-layer coupling between the CB and VB in each layer  as a 
perturbation that enters $S^{(1)}$. Following the same steps as for the 3R stacking, one finds 
that in the CB the Berry curvature is given by $\Omega_{z,cb}=\Omega_{z,cb}^{(0)}+\Omega_{z,cb}^{(1,1)}$, where
\begin{subequations}
\begin{equation}
 \Omega_{z,cb}^{(0)}(\mathbf{q})=\mp \frac{\tau}{2} 
  \frac{\gamma_{cc}^2 U_g}{\left(U_g^2+\left({\gamma_{cc} |\mathbf{q}|}\right)^2\right)^{3/2}}\\
 \label{eq:omega0-cb-2H}
\end{equation}
is due to the inter-layer coupling of the CBs.  
The second contribution reads 
\begin{equation}
\Omega_{z,cb}^{(1,1)}(\mathbf{q})\approx\pm\frac{\tau}{2} \left(\frac{\gamma_3}{\delta E_{bg}}\right)^2 
   \lambda_3 
  \frac{U_g}{\left(U_g^2+\left({\gamma_{cc} |\mathbf{q}|}\right)^2\right)^{1/2}}, 
 \label{eq:omega11-cb-2H}
\end{equation}
\label{eq:omega-cb-2H}
\end{subequations}
where, using the notation $\tilde{\varepsilon}_{vb}=\sqrt{t_{\perp}^2+U_g^2}$, 
the constant $\lambda_3$ is given by 
$\lambda_3=1+\frac{3}{4}\left(\frac{\tilde{\varepsilon}_{vb}}{\delta E_{bg}}\right)^2$ 
and terms $\sim \mathbf{q}^2$ have been neglected in Eq.~(\ref{eq:omega11-cb-2H}).
$\Omega_{z,cb}^{(1,1)}(\mathbf{q})$ is non-zero even if we set $\gamma_{cc}=0$, i.e., 
this term describes a Berry curvature contribution due to the intra-layer 
coupling of the CB and the VB. For the VB one finds that $\Omega_{z,vb}^{(0)}=0$ and the 
first non-zero term is 
\begin{equation}
\Omega_{z,vb}^{(1,1)}=\mp 2 \tau \frac{\gamma_3^2 U_g}{\tilde{\varepsilon}_{vb}(E_{bg}\mp\tilde{\varepsilon}_{vb})^2},
 \label{eq:omega11-vb-2H}
\end{equation}
which is  in agreement with  Ref.~\onlinecite{symmetry-tuning} for $\tilde{\varepsilon}_{vb}\ll E_{bg} $. 
This means that the Berry curvature is, in first approximation, 
dispersionless in the VB.
The upper (lower) sign in Eqs.~(\ref{eq:omega0-cb-2H})-(\ref{eq:omega11-cb-2H}) and (\ref{eq:omega11-vb-2H})
corresponds to the bands that have larger weight in the layer at $+U_g$ ($-U_g$) potential.  
One can note that the inter-layer ($\Omega_{z,cb}^{(0)}$) and intra-layer ($\Omega_{z,cb}^{(1,1)}$) contributions
have opposite sign in each valley. 
As shown in Figures~\ref{fig:Berry-curv-calc}(a) and (b), our numerical calculations using  the 
eigenstates of Eq.~(\ref{eq:H-3R-4dim}) and (\ref{eq:H-2H-4dim})
are in good agreement with the analytical results of Eqs.~(\ref{eq:omegaz-3r}) and 
Eqs.~(\ref{eq:omega-cb-2H})-(\ref{eq:omega11-vb-2H}).

\subsection{Discussion}

\begin{figure}[htb]
  \includegraphics[scale=0.45]{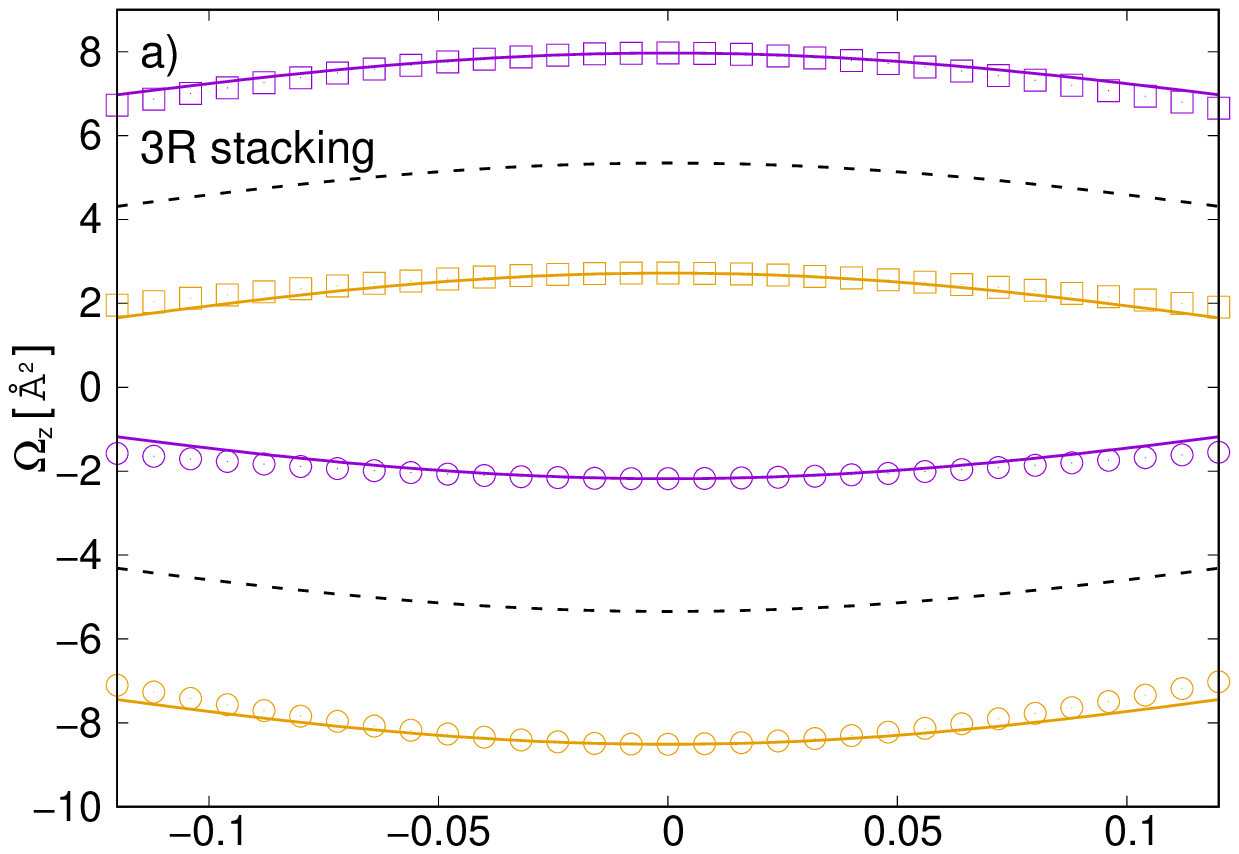}
  \includegraphics[scale=0.45]{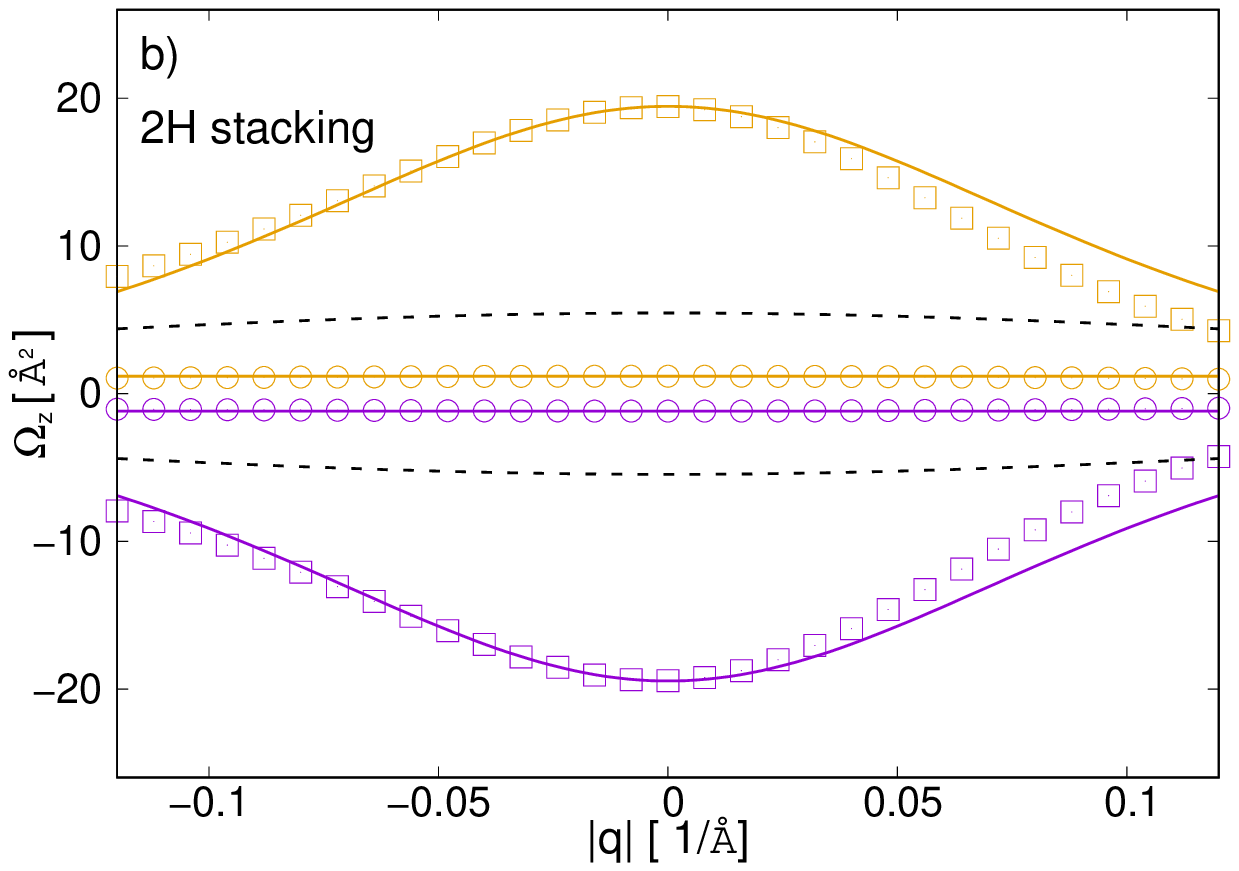}\\
  \includegraphics[scale=0.2]{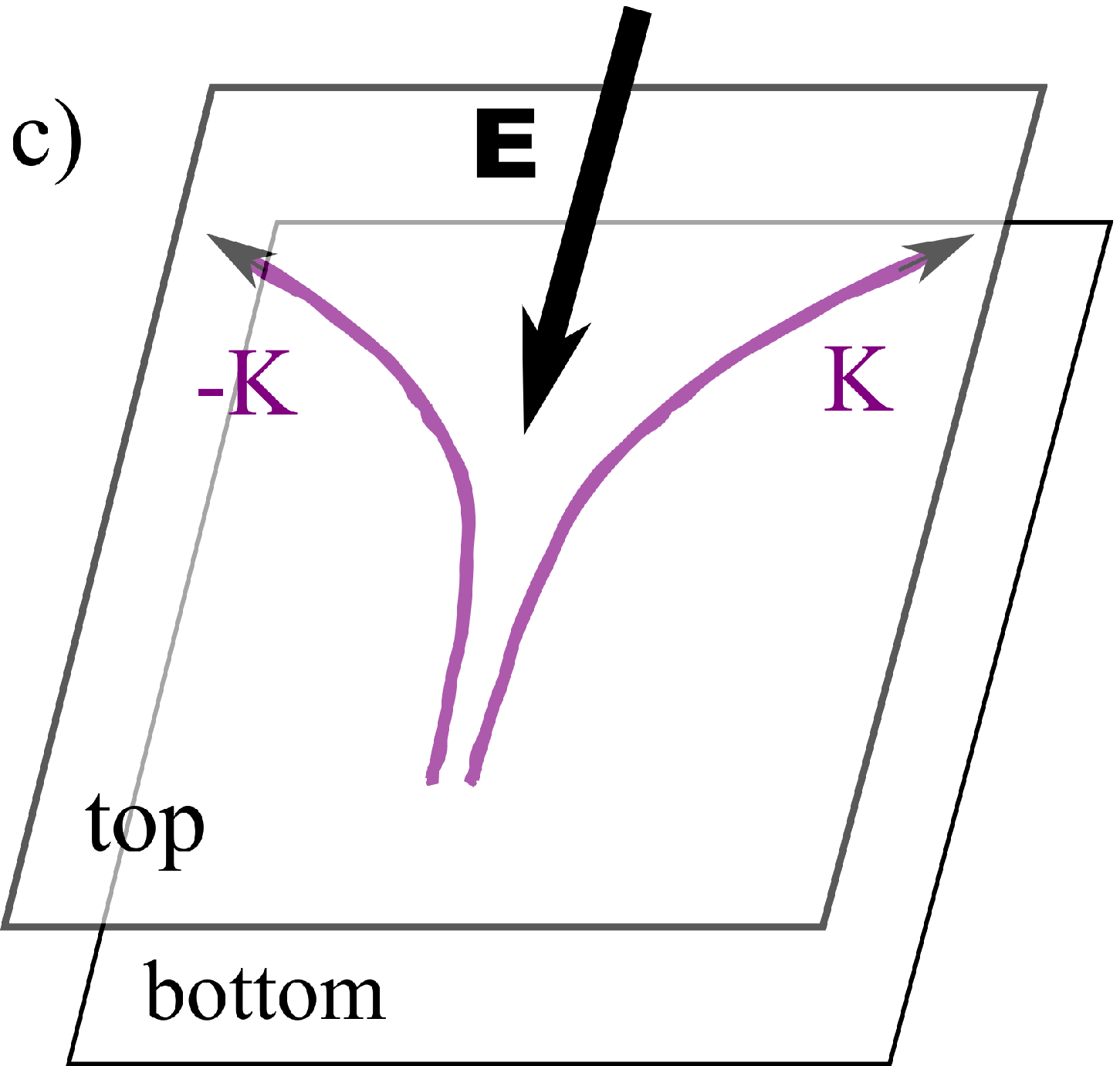} 
  \includegraphics[scale=0.2]{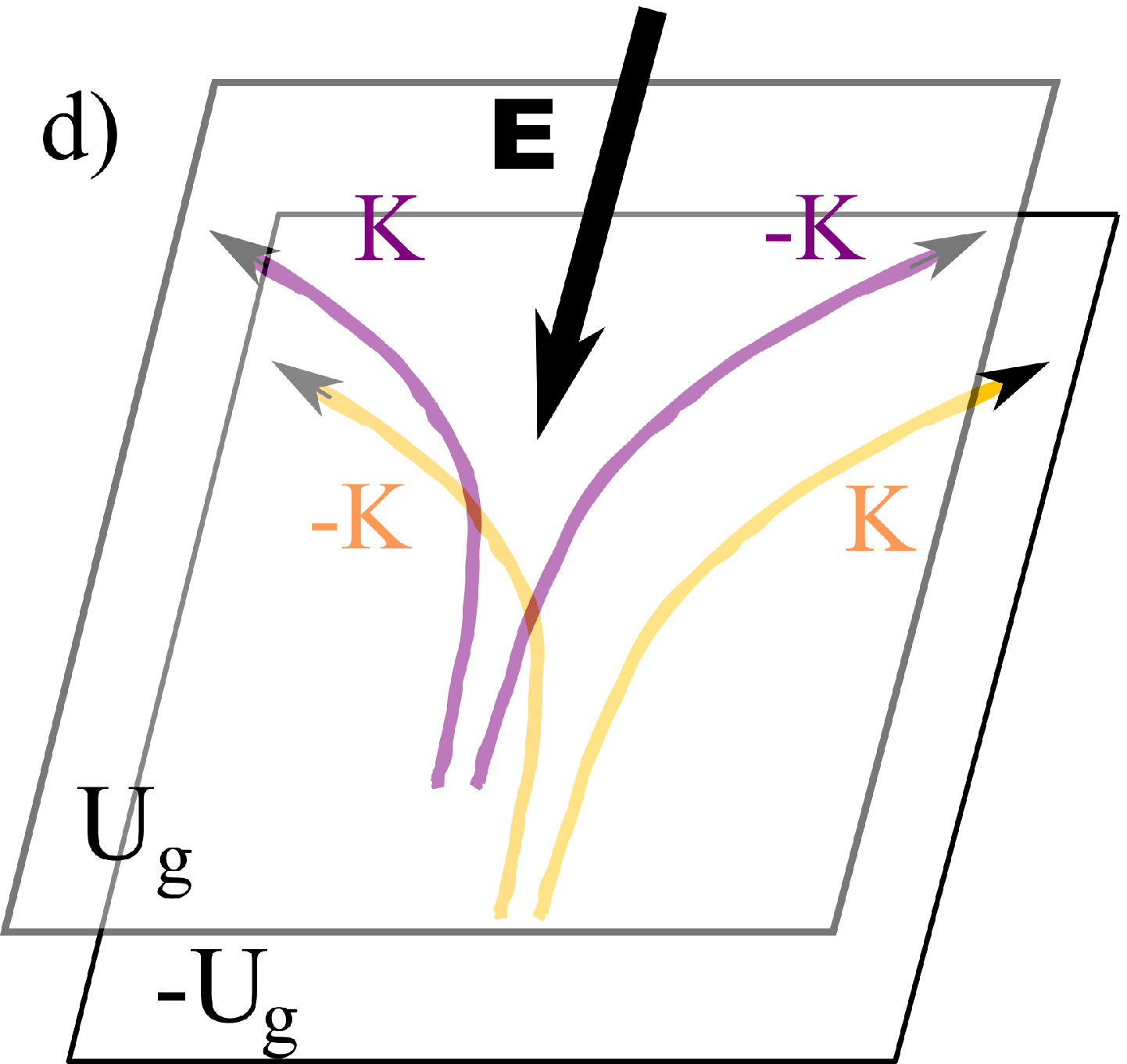}
 \caption{ {Comparison of numerical and analytical calculation of $\Omega_z$  around the $K$ point 
            for a) 3R stacked, and b) 2H stacked bilayer MoS$_2$. $\square$ show the results for CBs, $\bigcirc$ for  VBs.  
            In a),  brown colour corresponds to bands in the bottom layer, purple to bands in the top layer, 
            solid lines show the results of Eq.~(\ref{eq:omegaz-3r}).  
            Dashed lines indicate the Berry curvature $\Omega_z^{(0)}$ of a monolayer, given by Eq.~(\ref{eq:omega0-aa}).
            In b) brown colour corresponds to the layer at $-U_g$, purple to the layer at $+U_g$ potential, 
            solid lines show the results of 
           Eqs.~(\ref{eq:omega0-cb-2H})-(\ref{eq:omega11-cb-2H}) and (\ref{eq:omega11-vb-2H}), 
           dashed lines indicate  $\Omega_{z,cb}^{(0)}$ for the inter-layer 
           contribution given by Eq.~(\ref{eq:omega0-cb-2H}). For material parameters of the $\mathbf{k}\cdot\mathbf{p}$ 
           models see Table~\ref{tbl:params}. In b) we used $U_g=10$meV. The plotted range corresponds to around 
           10\% of the $\Gamma-K$ distance in the BZ. c) and d): Schematics of the valley Hall conductivity 
           contributions in the CB of  3R and 2H bilayers, respectively, when an in-plane electric field $\mathbf{E}$ is 
           applied.}  
\label{fig:Berry-curv-calc}}
\end{figure}

One can see that although the band structure of 3R  and 2H stacked bilayers look rather similar, 
especially in the valence band 
[c.f. Figures~\ref{fig:cryst-struct-bilayer}(a) and \ref{fig:cryst-struct-bilayer}(b)],
the comparison of Figs.~\ref{fig:Berry-curv-calc}(a) and \ref{fig:Berry-curv-calc}(b) 
reveals  several important differences between their Berry curvature properties. 
Considering first the 3R bilayers, the Berry curvature is essentially  \emph{layer-coupled} 
both in the VB and in the CB:  it is significantly larger  in the CB of the top layer than 
of the bottom layer, while  the converse is true for the VBs [see Fig.~\ref{fig:Berry-curv-calc}(a)]. 
In the CB of the bottom and top layers one finds 
for $\mathbf{q}=0$ that $\Omega_{z,cb}= \Omega_{z,cb}^{(0)} +\Omega_{z,cb}^{(1,1)}=
\frac{\tau}{2}[(\gamma_3/\delta E_{bg})^2 \mp (\gamma_{cc}/\delta E_{ll})^2]$, where 
$-$ $(+)$ sign is for the bottom (top) layer. This expression shows that i) both intra-layer 
and inter-layer coupling contribute to the Berry curvature, 
and (ii) the two contributions  can either reinforce or weaken each other. 
The effect of the inter-layer coupling is clearly visible: it reduces $\Omega_{z,cb}^{}$ 
for the bottom layer and enhances it for the top layer.  
A similar  but opposite effect  takes place in the VB as well, where 
$\Omega_{z,vb}=
-\frac{\tau}{2}[(\gamma_3/\delta E_{bg})^2 \pm (\gamma_{vv}/\delta E_{ll})^2]
$. 
Using the band-structure parameters given in Table~\ref{tbl:params}, we find that 
the intra-layer   and the inter-layer contributions are of similar magnitude: 
although the coupling $\gamma_{cc}$ between the layers is much weaker 
than the intra-layer coupling $\gamma_{3}$ between the CB and the VB, 
since $\delta E_{ll}\ll \delta E_{bg}$,  the ratios $\gamma_3/\delta E_{bg}$ 
and $\gamma_{cc}/\delta E_{ll}$ are of the same order of magnitude. 
This conclusion  does not seem to depend on the level of theory applied 
in the first-principles calculations 
which yield the band-structure  parameters for the $\mathbf{k}\cdot\mathbf{p}$ theory: 
using, as an estimate, the parametrization for $\gamma_3$ and $\delta E_{bg}$ 
of e.g., Ref.~\onlinecite{Kaxiras} 
which are based on GW calculations for the 2H polytype,  one still finds that 
the ratio $\gamma_3/\delta E_{bg}$ is similar in the DFT and GW calculations and 
we expect the same for  $\gamma_{cc}/\delta E_{ll}$. 
Moreover,  $\Omega_{z}^{(1,1)}$ and hence the total Berry curvature may be \emph{tunable} by an 
external electric field which would  change  $\delta E_{ll}$. 
We point out that the external electric field can,  in principle,  both \emph{decrease} and \emph{increase}   
$\delta E_{ll}$ depending on its polarity, and the same can be expected for $\Omega_{z}^{(1,1)}$ as well
(assuming that $\gamma_{cc}$ and $\gamma_{vv}$ do not change significantly). While $E_{bg}$ and hence 
$\Omega_{z}^{(0)}$ is difficult to change by electric field because it depends on the crystal field splitting
of the atomic Mo $d$ orbitals, $\delta E_{ll}$ is determined
by weaker inter-layer interactions and hence it  might be more easily tunable. 

For 2H bilayers, on the other hand, the Berry curvature is \emph{CB-coupled}: 
it is much larger in the CB than in the VB [see Fig.~\ref{fig:Berry-curv-calc}(b)]. 
This can be understood from Eqs.~(\ref{eq:omega0-cb-2H}) and (\ref{eq:omega11-cb-2H}): 
for small $\mathbf{q}$ values, such that $\gamma_{cc}|\mathbf{q}|<< U_g$
the main contribution to $\Omega_{z,cb}$ comes from the inter-layer term $\Omega_{z,cb}^{(0)}$
and can be quite large for small $U_g$ values. Similarly to 3R bilayers, therefore, 
$\Omega_{z,cb}$ is gate tunable. 
In contrast, using Eq.~(\ref{eq:omega11-vb-2H}) we expect that the Berry curvature, albeit 
gate tunable, will be rather small in the VB. Assuming  $U_g$  of the order of $1-10$\,meV 
which  we think is experimentally feasible,  one finds that 
$U_g$ is significantly  smaller than $t_{\perp}$ (see Table \ref{tbl:params}) and therefore  
$\Omega_{z,vb}^{(1,1)} \sim 0.5 - 1.0$\,\AA$^2$.

\begin{table}[htb]
\begin{tabular}{|c|c|c|c|c|c|}\hline
   & $t_{\perp}$ [eV] & $\gamma_{cc}$ [eV\AA] & $\gamma_{vv}$ [eV\AA] &  $\delta E_{cc}$ [eV] & $\delta E_{vv}$ [eV]\\
 \hline
3R &  -      & $0.0708$ & $0.0779$  & $0.029$ & $0.033$ \\
2H & $0.045$ & $0.0706$ & -         &   -     & - \\
 \hline
\end{tabular}
\caption{Material parameters obtained by fitting the DFT band structure calculations which do not take into account
SOC, using Eqs.~(\ref{eq:H-3R-4dim}) and  (\ref{eq:H-2H-4dim}). 
To obtain Figs.~\ref{fig:Berry-curv-calc}(a) and (b) we used $\gamma_3=2.73$\,eV\AA\, and $E_{bg}=1.67$eV 
from Ref.~\onlinecite{2DMaterials-paper} and $\delta E_{ll}=0.031$eV.}
\label{tbl:params}
\end{table}


\section{Spin-orbit coupling effects}
\label{sec:SOC}

The considerations in Sections \ref{sec:kp-Ham} and \ref{sec:Berry-curv} should be applicable to
all homobilayer TMDCs. We have not yet discussed the effect of the spin-orbit coupling (SOC) on the 
Berry curvature properties of BTMDCs. Generally, the SOC in BTMDCs is more complex
than in monolayers, see the Appendices \ref{subsubsec:SOC-K-pt-2H} and \ref{subsubsec:SOC-K-pt-3R} for details. 
Moreover,  in 3R bilayers the low-energy physics also depends on the ratio of the 
band-edge energy difference $\delta E_{cc}$ and $\delta E_{vv}$ and the monolayer SOC coupling strengths
$\Delta_{cb}$ and $\Delta_{vb}$. These energy scales can be quite  different in different BTMDCs. 
Because of the recent experimental activity\cite{valley-Hall-BLMoS2,Morpurgo-3R-bilayer} 
we will  focus on bilayer MoS$_2$. 
Our DFT calculations suggest that for bilayer MoS$_2$  it is sufficient to take into account only the 
intrinsic SOC of the constituent monolayers.   

\subsection{3R bilayer MoS$_2$}

\begin{figure*}[htb] 
   \includegraphics[scale=0.65]{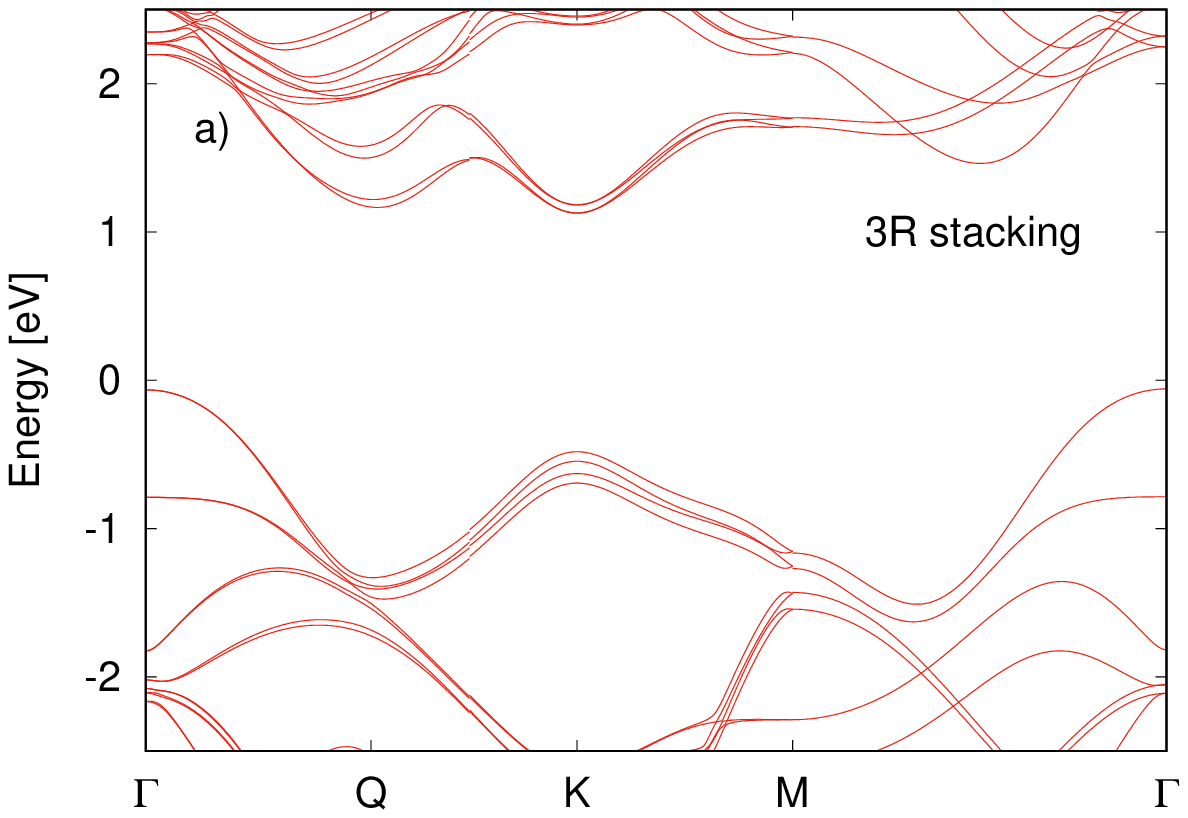}
    \includegraphics[scale=0.65]{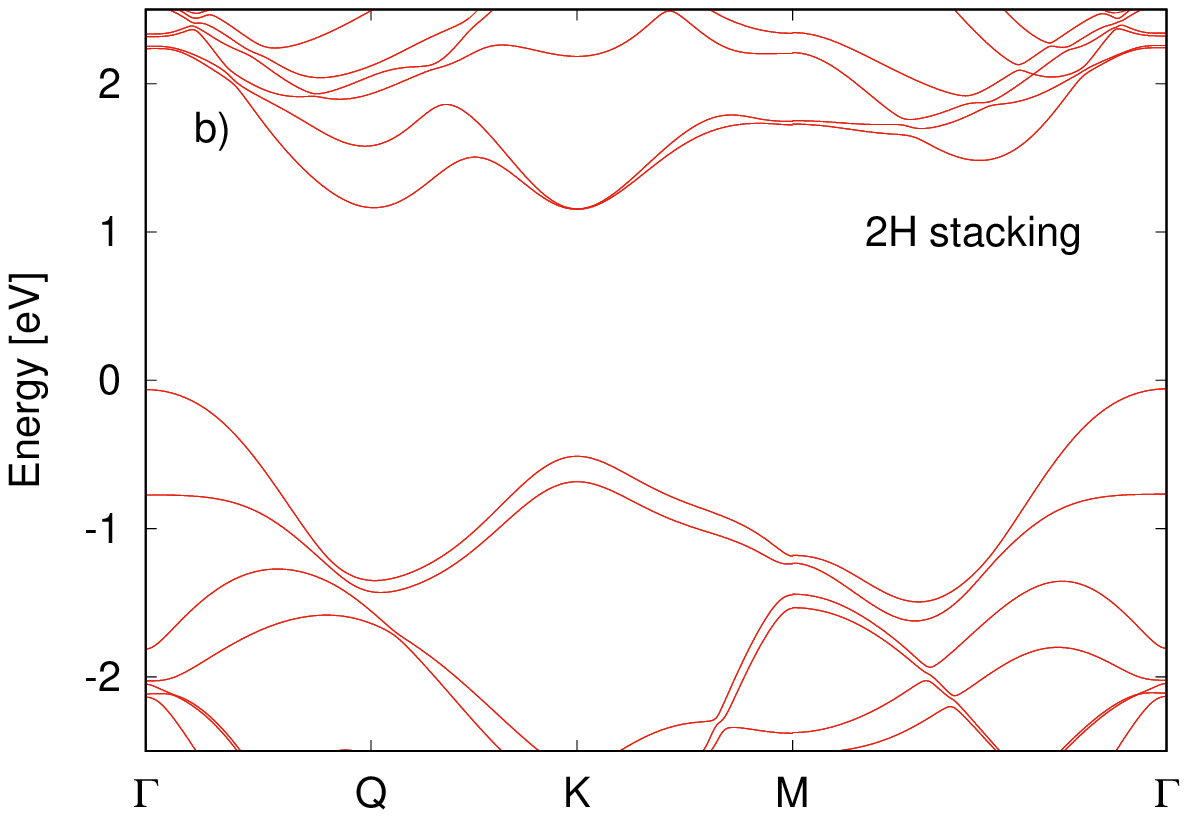}
 \includegraphics[scale=0.5]{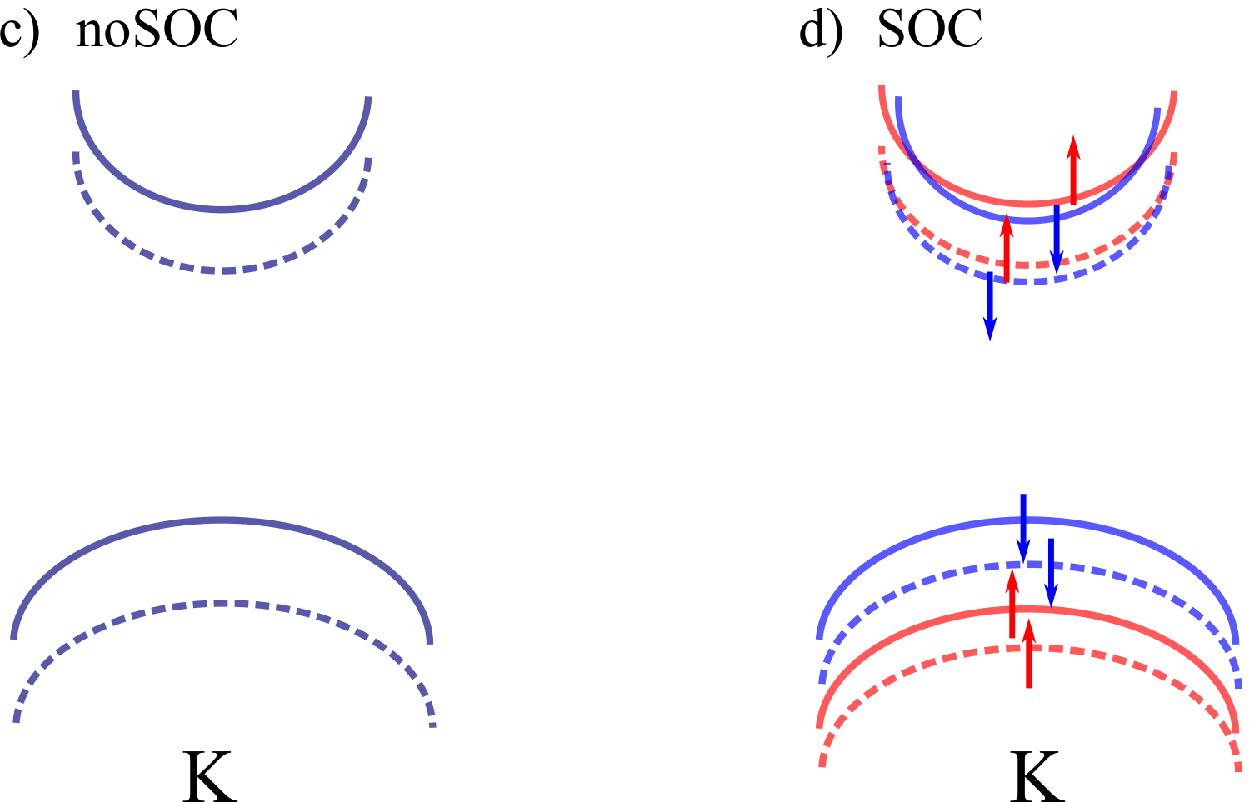}\hspace{1.5cm}
  \includegraphics[scale=0.5]{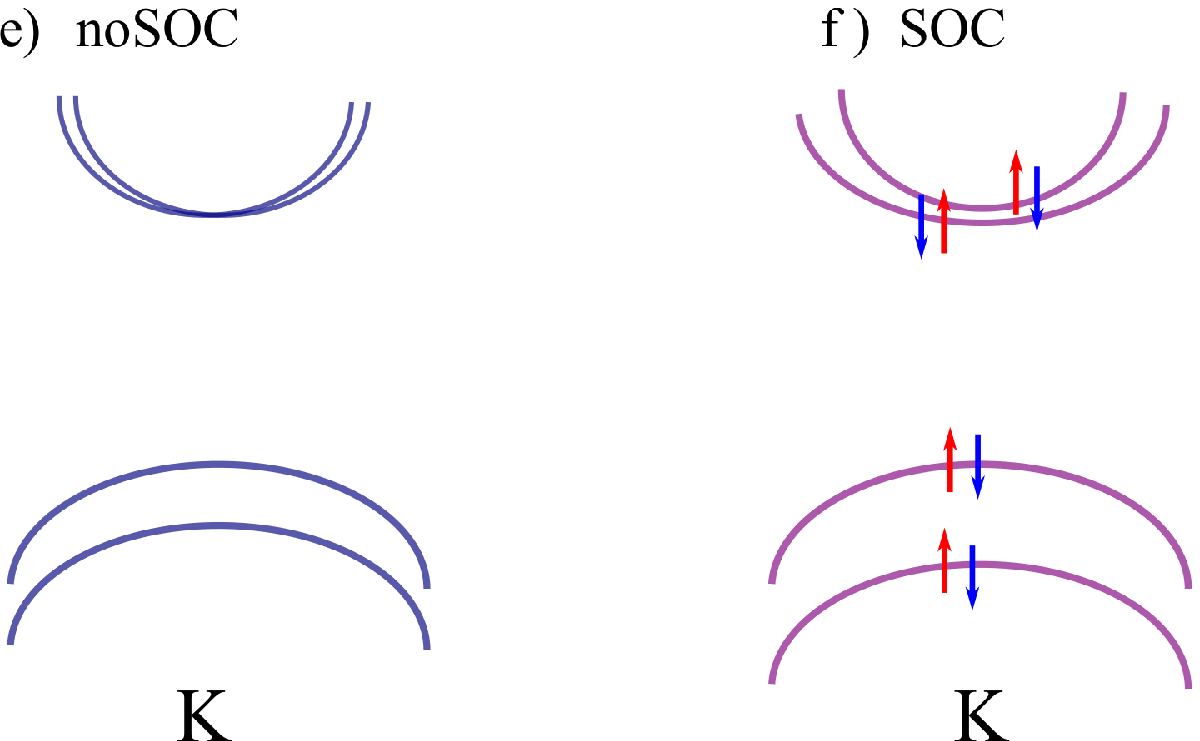}
  \caption{{SOC effects in the band structure of bilayer MoS$_2$.  
       a) DFT band structure calculations along the $\Gamma-K-M-\Gamma$ line of the BZ for 3R bilayer.
       b) The same as in a) for 2H bilayer.
       c) For 3R bilayers, close to the $\pm K$ points  the layer index is an approximately 
       good quantum number  for each of the bands  both in the CB and the VB. 
       Neglecting the SOC (noSOC) the lowest CB is mostly localized to the top layer (dashed line), 
       while the next CB band (solid line) to the bottom layer. The opposite is true for the two highest 
       energy VBs. The bands are shifted in energy due to the inter-layer band edge energy difference 
       $2\delta E_{cc}$ and $2\delta E_{vv}$. 
       d) When SOC is taken into account for 3R bilayers, each of the bands becomes spin-split and spin polarized. 
       Red corresponds to $\uparrow$, blue to $\downarrow$ spin polarization. 
       The spin-splitting $\Delta_{cb}$ of the two lowest CBs is much 
       smaller than the inter-layer splitting $\delta E_{cc}$. The situation is different for the VBs: 
       here $\Delta_{vb}\gtrsim \delta E_{vv}$ and therefore the spin-polarized bands have an alternating 
       layer index. 
       e) For 2H bilayers, if SOC is neglected (noSOC), the two lowest energy CB are degenerate 
          at the $\pm K$ points and weakly split due to the inter-layer coupling away from the $\pm K$ points. 
          The energy splitting of the two highest energy VBs is $2 t_{\perp}$. Both layers contribute with equal weight 
          to each of the bands. 
       f) When SOC is taken into account for 2H bilayers, in the CB there are two two-fold degenerate 
          and spin-unpolarized bands separated by an energy  $2\Delta_{cb}$ at the $\pm K$ point. 
          A combined layer and spin index  can be assigned to each of the four CB bands at the $\pm K$ point,  
          away from  the $\pm K$ points both layers contribute to each of the bands, but with different weights. 
          In the VB both layers contribute to each of the bands, even at 
          the $\pm K$ points. Only if $\Delta_{vb}\gg t_{\perp}$ do the bands become approximately layer 
          polarized\cite{WangYao-bilayer}. 
          In d) and f) the spin-polarization of the bands in the $-K$ valley can be obtained by taking 
          the time reversed states. }
 \label{fig:SOC-bands-bilayer}}
 \end{figure*}

Fig.~\ref{fig:SOC-bands-bilayer}(a) shows  the band structure of 3R bilayer MoS$_2$ obtained from DFT calculations. 
In contrast to Fig.~\ref{fig:cryst-struct-bilayer}(a)  here the SOC is also taken into account. 
The effects of the SOC at the $K$ point of the BZ are highlighted by comparing the  schematic band structure without 
and with the SOC in Figs.~\ref{fig:SOC-bands-bilayer}(c) and \ref{fig:SOC-bands-bilayer}(d), respectively.
As already explained in Section \ref{sec:kp-Ham}, the band edge energy is different for 
bands localized to the top and bottom layers both in the CB and the VB. Upon considering the SOC, 
since 3R bilayers lack inversion symmetry, their bands, apart from the $\Gamma-M$ direction in the BZ, 
will be spin-orbit split and spin-polarized [Fig.~\ref{fig:SOC-bands-bilayer}(a)].  The SOC is diagonal in the layer index
and it can be described by   adding  a term  $H_{so,cb}^{3R}=\Delta_{cb} \tau_z s_z$ ($H_{so,vb}^{3R}= \Delta_{vb} \tau_z s_z$) 
to the  CB  (VB) of the constituent monolayers, where 
$\Delta_{cb (vb)}$ is (in good approximation) the SOC strength 
in  monolayer MoS$_2$, $s_z$ is a spin Pauli matrix and the Pauli matrix $\tau_z$ acts in the valley space. 
Our DFT calculations show  that  $\Delta_{cb}$ is much smaller than the inter-layer band-edge energy 
difference $\delta E_{cc}$.  Therefore, as shown schematically in Fig.~\ref{fig:SOC-bands-bilayer}(d), 
it has a minor effect on the band structure.  The situation is different in the VB, because 
$\Delta_{vb}$ is larger than the band-edge energy difference $\delta E_{vv}$. 
Therefore in the $\pm K$ valleys the four highest energy spin-split VB show an alternating 
layer polarization pattern.

Regarding the Berry curvature calculations, the effect of SOC on 
the formulas  Eqs.~(\ref{eq:omegaz-3r}) can be rather straightforwardly taken into account by introducing 
the spin-dependent band gaps $\delta E_{bg}^{\downarrow}=\delta E_{bg}-\tau (\Delta_{cb}+\Delta_{vb})/2$ 
and  $\delta E_{bg}^{\uparrow}=\delta E_{bg}+\tau (\Delta_{cb}+\Delta_{vb})/2$ and the corresponding 
spin-dependent Berry curvatures for the top and bottom layers. Note that $\delta E_{ll}$ 
in Eq.~(\ref{eq:omega11a-aa}) \emph{is not affected} by the SOC because in 3R bilayers, unlike
in 2H bilayers, the SOC is not layer dependent and therefore it drops out from the 
inter-layer energy difference. 


\subsection{2H bilayer MoS$_2$}

Turning now to  the 2H bilayers, we remind that 
if  SOC is neglected and inversion symmetry is not broken, the CB is doubly degenerate, 
while the VB is non-degenerate in the $\pm K$ point [Fig.~\ref{fig:cryst-struct-bilayer}(b)]. 
If now spin is taken into account but SOC is neglected, this would mean a 
four-fold degeneracy of the CB.  However, the SOC partially lifts this four-fold degeneracy 
and leads to two two-fold degenerate levels, see Figs.~\ref{fig:SOC-bands-bilayer}(e) and \ref{fig:SOC-bands-bilayer}(f). 
In contrast to the 3R bilayers, due to the inversion symmetry
all bands of 2H bilayers remain spin-degenerate even when SOC is considered.
The SOC of 2H bilayers  can be described by  the Hamiltonian  $H_{so,cb}^{2H}=\Delta_{cb} \tau_z\sigma_z s_z$ 
($H_{so,vb}^{2H}=\Delta_{vb} \tau_z \sigma_z s_z$) in the CB (VB) of the
bilayer. Here the Pauli matrix $\sigma_z$ indicates that within a given valley 
the SOC has a different sign\cite{WangYao-bilayer} 
in the two layers: this can be understood from the fact that the layers are rotated by $180^{\circ}$ with 
respect to each other. At each energy there will be a $\uparrow$ and a $\downarrow$ polarized band, 
see Fig.~\ref{fig:SOC-bands-bilayer}(f). In the CB  the splitting between the two-fold degenerate 
levels is essentially given by the SOC strength $2 \Delta_{cb}$ of monolayer TMDCs. 
In the VB the main effect of the SOC is to increase the energy splitting of the two highest bands 
from $2 t_{\perp}$ to  $2 \sqrt{\Delta_{vb}^2+t_{\perp}^2}$.

The SOC has  an interesting effect on the Berry curvature. 
Considering first the CB, the SOC  leads to a finite $\Omega_{z,cb}^{(0)}$ even for $U_g=0$, i.e., when 
there is no external electric field applied. 
The corresponding formulas can be obtained from Eqs.~(\ref{eq:omega0-cb-2H}), (\ref{eq:omega11-cb-2H}) 
by making the substitution $U_g\rightarrow \Delta_{cb}$  
and using  $\tilde{\varepsilon}_{vb}=\sqrt{\Delta_{vb}^2+t_{\perp}^2}$ in the expression for $\tilde{\lambda}_3$.
One can label $\Omega_{z}$ by a spin index  $s=\uparrow,\downarrow$ and write 
$\Omega_{z,cb}^{\uparrow}=\Omega_{z,cb}^{(0)}+\Omega_{z,cb}^{(1,1)}$,  where 
the upper (lower) sign appearing in Eqs.~(\ref{eq:omega0-cb-2H}) and (\ref{eq:omega11-cb-2H}) 
corresponds to the band at energy $\varepsilon_{cb}+\tau\Delta_{cb}$ ($\varepsilon_{cb}-\tau\Delta_{cb}$) for $\mathbf{q}=0$. 
Regarding the $\downarrow$ bands, one finds  $\Omega_{z,cb}^{\downarrow}=-\Omega_{z,cb}^{\uparrow}$.
In the VB, for the physically relevant spin-degenerate band at the band gap  one finds  
\begin{equation}
\Omega_{z,vb}^{(1,1,s)}=
\tau \cdot s   \frac{2 \gamma_3^2 \Delta_{vb}}{\tilde{\varepsilon}_{vb}(E_{bg}-\tilde{\varepsilon}_{vb})^2},
 \label{eq:omega11-vb-ab-SOC}
\end{equation}
where $s=1$ for the $\uparrow$ ($s=-1$ for $\downarrow$) spin-polarized band. This  result was also obtained in 
Ref.~\onlinecite{WangYao-bilayer}.


\section{Valley Hall effects}
\label{sec:valley-Hall}

We now discuss how the Berry curvature affects the valley and spin Hall conductivities 
in bilayer MoS$_2$. Since few-layer MoS$_2$ on dielectric substrate is often found to be
$n$-doped\cite{Tartakovskii}, we will focus on   the valley Hall effects in the CB. 
Although the Q and K valleys are nearly degenerate, in our DFT calculations the band edge in the CB 
is at the $\pm K$ point. 
Therefore we can use the results obtained in Sections \ref{sec:Berry-curv} and \ref{sec:SOC}. 
The relevance  of the $Q$ point valleys in the CB will be briefly discussed in Section~\ref{sec:Q-pt-main}. 
Regarding the VB, we briefly note that the band edge energy difference $E_{\Gamma K}$ 
between the $\Gamma$ and $K$ points is quite large, $500-600$\,meV and therefore 
the $K$ valley is not relevant for transport properties of $p$-doped samples. 
Nevertheless, in other BTMDCs $E_{\Gamma K}$  might be much smaller\cite{Franchini,Mauri} 
than in MoS$_2$ and therefore the $\pm K$ valleys may also play an important  role. 
We leave the study of the valley Hall effect in the VB of BTMDCs to a future work.

Due to the Berry curvature, if an in-plane electric field is applied, the charge carriers will acquire  a 
transverse anomalous velocity component\cite{Berry-phase-review} which gives rise to an 
intrinsic contribution to the Hall conductivity\cite{AnomalousHall-review}. 
We may  define the valley Hall conductivity  $\sigma_{v,H}^{}$ of band $n$ as\cite{WangYao-TMDCmonolayer,AnomalousHall-review} 
\begin{equation}
 \sigma_{n,v,H}^{}=\frac{e^2}{\hbar}\int\frac{d \mathbf{q}}{(2\pi)^2} 
  \left[ f_n^{\uparrow}(\mathbf{q}) \Omega_{z,n}^{\uparrow}(\mathbf{q})+
  f_n^{\downarrow}(\mathbf{q}) \Omega_{z,n}^{\downarrow}(\mathbf{q})\right]
 \label{eq:sigmavH}
\end{equation}
where  $f_n^{\uparrow,\downarrow}(\mathbf{q})$ is the Fermi-Dirac 
distribution function. 
Similarly, the spin Hall conductivity can be defined as 
\begin{equation}
 \sigma_{n,s,H}^{}=\int\frac{d \mathbf{q}}{(2\pi)^2}\left[ 
  f_n^{\uparrow}(\mathbf{q}) \Omega_{z,n}^{\uparrow}(\mathbf{q})-
  f_n^{\downarrow}(\mathbf{q}) \Omega_{z,n}^{\downarrow}(\mathbf{q})\right].
 \label{eq:sigmasH}
\end{equation}
Since we only study the valley Hall effects in the CB, we neglect the band index $n$ in the following. 
For later reference we note that since for each band one may write the Berry curvature 
as $\Omega_{z}=\Omega_z^{(0)}+\Omega_{z}^{(1,1)}$, the corresponding conductivities   
read  $\sigma_{v,H}^{}=\sigma_{v,H}^{(0)}+\sigma_{v,H}^{(1,1)}$ and 
$\sigma_{s,H}^{}=\sigma_{s,H}^{(0)}+\sigma_{s,H}^{(1,1)}$.

\subsection{3R bilayers}
Due to the relatively large band-edge energy difference $2 \delta E_{ll}=58$\,meV,    
for typical $n$-doping only the  CB (mostly) localized to the top layer would be 
occupied and  have a finite $\sigma_{v,H}$ and $\sigma_{s,H}$ contribution 
[the former is shown schematically  in Fig.~\ref{fig:Berry-curv-calc}(c)].
This situation is similar to one of the  proposed strongly interacting phases of bilayer graphene, namely, 
to the Quantum Valley Hall insulator phase\cite{BLG-Hallstates}.   However, in our case $\sigma_{v,H}$ is not  quantized. 
In the following we assume that the charge density is large enough so that both spin-split CB bands 
in the top layer are populated and we add up their contributions to $\sigma_{v,H}$ and $\sigma_{s,H}$. 
Since $\Omega_{z}(\mathbf{q})$ changes rather slowly around the 
$\pm K$ points [see Fig.~\ref{fig:Berry-curv-calc}(a)] we may use $\Omega_{z}(\mathbf{q}=0)$ 
in Eq.~(\ref{eq:sigmavH}) and at zero temperature one finds   
\begin{eqnarray}
\sigma_{v,H}^{3R}&=\frac{\tau}{2} \frac{e^2}{\hbar}
  \left[
  \left(\frac{\gamma_{cc}}{\delta E_{cc}}\right)^2
   \frac{q_{F,\downarrow}^2+q_{F,\uparrow}^2}{4\pi}  
   \right . \nonumber\\
   &+
   \left. 
   \left(\frac{\gamma_3}{\delta E_{bg}^{\uparrow}}\right)^2 \frac{q_{F,\uparrow}^2}{4\pi}
   +\left(\frac{\gamma_3}{\delta E_{bg}^{\downarrow}}\right)^2 \frac{q_{F,\downarrow}^2}{4\pi}   
   \right]      
   \label{eq:sigmavH-3R}
\end{eqnarray}
where $q_{F,\downarrow}$  ($q_{F,\uparrow}$) is the Fermi wavevector for electrons of  $\downarrow$ ($\uparrow$) spin.
The  term in the first line on the  right-hand-side (r.h.s) of Eq.~(\ref{eq:sigmavH-3R})   corresponds 
to $\sigma_{v,H}^{(1,1)}$ while the second line is $\sigma_{v,H}^{(0)}$. One may recognize that $\sigma_{v,H}^{(0)}$
equals  the valley Hall conductivity $\sigma_{v,H}^{ml}$ of monolayer TMDCs\cite{WangYao-TMDCmonolayer}. 
Note that $(q_{F,\downarrow}^2+q_{F,\uparrow}^2)/{4\pi}=n_{e,v}$ is the total charge density per valley.  After expanding 
the second line  of Eq.~(\ref{eq:sigmavH-3R}) in terms $(\Delta_{cb}+\Delta_{vb})/(2\delta E_{bg})$, one finds that it 
is also $\sim n_{e,v}$ plus a correction $\sim (q_{F,\uparrow}^2-q_{F,\downarrow}^2)$ which is typically small with 
respect to terms that are $\sim n_{e,v}$. 
{Since $\gamma_3/\delta E_{bg}$ and $\gamma_{cc}/\delta E_{ll}$ are of the same order of magnitude, Eq.~(\ref{eq:sigmavH-3R})
shows that $\sigma_{v,H}^{3R}$ is roughly twice as big as $\sigma_{v,H}^{ml}$.}
The sign of $\sigma_{v,H}^{3R}$ is opposite in the $K$ and $-K$ valleys, therefore no net bulk 
charge current  flows   unless there is a charge imbalance between the valleys. 
Calculating the spin Hall conductivity e.g., in the $K$ valley, it reads 
\begin{eqnarray}
  \sigma_{s,H}^{3R}&=\frac{1}{2} 
  \left[
  \left(\frac{\gamma_{cc}}{\delta E_{cc}}\right)^2
   \frac{q_{F,\uparrow}^2-q_{F,\downarrow}^2}{4\pi}  
   \right . \nonumber\\
   &+
   \left. 
   \left(\frac{\gamma_3}{\delta E_{bg}^{\uparrow}}\right)^2 \frac{q_{F,\uparrow}^2}{4\pi}
   -\left(\frac{\gamma_3}{\delta E_{bg}^{\downarrow}}\right)^2 \frac{q_{F,\downarrow}^2}{4\pi}   
   \right].
   \label{eq:sigmasH-3R}
\end{eqnarray}
The second line in Eq.~(\ref{eq:sigmasH-3R}), which corresponds to $\sigma_{s,H}^{(0)}$, 
is the same as in monolayer MoS$_2$.  Because of the $\sigma_{s,H}^{(1,1)}$
contribution shown in the first line on the r.h.s of  Eq.~(\ref{eq:sigmasH-3R}), 
 $\sigma_{s,H}^{3R}$  is larger than $\sigma_{s,H}^{ml}$ in monolayer MoS$_2$. 
The term $q_{F,\uparrow}^2-q_{F,\downarrow}^2$ can be expressed 
as $(q_{F,\uparrow}^2-q_{F,\downarrow}^2){}= \frac{2 (m_{\uparrow}-m_{\downarrow})}{\hbar^2} E_F + 
\frac{2 m_{\uparrow}}{\hbar^2}\Delta_{cb}$ where $m_{\uparrow}$ and $m_{\downarrow}$ 
are the effective masses of the spin-split bands. Therefore the enhancement of $\sigma_{s,H}^{3R}$ 
depends on the Fermi energy $E_F$ and  on $\Delta_{cb}$. Our DFT calculations suggest that in MoS$_2$  
the term $\frac{2 m_{\uparrow}}{\hbar^2}\Delta_{cb}$ would dominate for $E_F\lesssim$ few tens of meV 
because $m_{\uparrow}-m_{\downarrow}\approx 0.03 m_e$ is rather small (here $m_e$ is the free electron mass). 
As we discussed for $\sigma_{v,H}^{3R}$,  $\sigma_{s,H}^{(0)}$ can be expanded 
in terms of $(\Delta_{cb}+\Delta_{vb})/(2\delta E_{bg})$ and we find that $\sigma_{s,H}^{3R}$ is 
roughly twice as big as $\sigma_{s,H}^{ml}$.  One can also easily show that,  
as in monolayers\cite{WangYao-TMDCmonolayer}, the magnitude and sign of $\sigma_{s,H}^{3R}$ does not depend 
on the valley index $\tau$.


\subsection{2H bilayers}

The situation is more complex for 2H bilayers than for their 3R counterparts.
As a first step we will discuss the valley Hall and spin  Hall effects qualitatively. 
Let us start with the $U_g=0$ case. 
As already mentioned in Section \ref{sec:SOC},  the SOC  leads to a finite  Berry curvature even for $U_g=0$.
Since inversion symmetry is not broken and therefore each band is spin-degenerate, 
 $f_n^{\downarrow}(\mathbf{q})= f_n^{\uparrow}(\mathbf{q})$.  On the other hand,   one finds that 
 $\Omega_{z}^{\uparrow}(\mathbf{q})=-\Omega_{z}^{\downarrow}(\mathbf{q})$ and therefore 
 $\sigma_{v,H}^{2H}$ vanishes in this limit. 
 However, $\Omega_{z}^{\uparrow}(\mathbf{q})-\Omega_{z}^{\downarrow}(\mathbf{q})$, and hence     
 $\sigma_{s,H}$ are non-zero. This is allowed because both the 
 (in-plane) electric field and the spin current transform in the same way under  
 time-reversal and inversion symmetries\cite{Murakami}.

\begin{figure}[htb] 
  \includegraphics[scale=0.5]{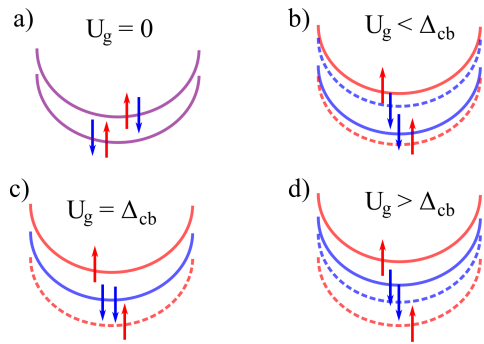}
  \caption{Schematic evolution of the four low-energy CB bands as a function of the inter-layer potential $U_g$ at 
  the $K$ point of the BZ. Spin-degenerate bands are shown with purple, $\uparrow$ polarized with red and 
  $\downarrow$ polarized with blue. Solid line corresponds to bands at  $+U_g$, dashed line to bands at $-U_g$  potential.  
 \label{fig:schem-Ug-2H}}
 \end{figure}

In general, for $U_g>0$ both  $\sigma_{v,H}^{2H}$ and  $\sigma_{s,H}^{2H}$ will be  finite.
For concreteness, we consider the $K$ point and first discuss qualitatively the evolution of the
band structure and the valley Hall conductivity as a function of $U_g$. 
The finite interlayer potential difference leads to the breaking of inversion symmetry and 
splitting of the spin-degenerate bands, as shown in  Figs.~\ref{fig:schem-Ug-2H}(a) and \ref{fig:schem-Ug-2H}(b). 
Each band can be labelled by a spin index $\uparrow$, $\downarrow$ and by the 
index $\pm$ depending on whether the band edge is  at $\pm U_g$ potential for $\mathbf{q}=0$. 
Next,  when $U_g=\Delta_{cb}$  [Fig.~\ref{fig:schem-Ug-2H}(c)]  the $(+,\downarrow)$ and $(-,\downarrow)$ 
bands become degenerate. We will show that upon further increasing $U_g$ [Fig.~\ref{fig:schem-Ug-2H}(d)], 
the contribution $\sigma_{v,H}^{(0)}$ ($\sigma_{s,H}^{(0)}$) to the total valley Hall (spin Hall) conductivity, 
which is due to the inter-layer coupling (see the discussion below Eq.~\ref{eq:omega0-cb-2H}), 
\emph{changes sign}. This behaviour is reminiscent of the topological transition in lattice 
Chern insulators\cite{Bernevig-book,AndrasPalyi}. 
Note however, that  the true band gap of the system, between the valence and conduction bands,  does not close. 
Nor does the gap close and re-open for the $(\uparrow,+)$ and $(\uparrow,-)$ bands. 
Therefore i) $\sigma_{v,H}^{2H}$ and $\sigma_{s,H}^{2H}$ are not quantized, and 
ii) those contributions to $\sigma_{v,H}^{2H}$ and $\sigma_{s,H}^{2H}$  
which are related to the intra-layer coupling of the CBs and the VBs 
do not change sign as a function of $U_g$.  
At the $-K$ point, by time reversal symmetry,  the $(\uparrow,+)$  and  $(\uparrow,-)$ bands can become 
degenerate as a function of  $U_g$.

We note that in recent experiments\cite{valley-Hall-BLMoS2,Morpurgo-3R-bilayer} 
the bilayer devices were fabricated with  a single backgate. 
In such devices the bilayer would be doped and at the same time a  finite inter-layer potential 
difference $U_g$ would be induced by changing the backgate voltage. 
Depending on the  $E_F/U_g$ and $E_F/\Delta_{cb}$ ratios, where $E_F$ is the Fermi energy, 
only one or both layers [this case  is illustrated  in Fig.~\ref{fig:Berry-curv-calc}(d)] and $1-4$ 
bands  may contribute to $\sigma_{v,H}^{2H}$ and $\sigma_{s,H}^{2H}$. 
In the following we will assume  that $E_F>2(U_g+\Delta_{cb})$ for all $U_g$ values considered, 
i.e.,  $E_F$  is large enough so that both layers and 
all four low-energy CBs  are occupied and contribute to the valley and spin Hall effects. 
In MoS$_2$, given the  relatively small $\Delta_{cb}\approx 3$meV value of the SOC,  we expect that this situation  
is realistic.  However,  in other 2H-BTMDCs where the SOC constant  $\Delta_{cb}$ can be
substantially larger than in MoS$_2$, not all four CBs would be necessarily occupied.  
Furthermore, we  neglect any Rashba type SOC induced by the external electric field because 
we expect that in the  devices of Refs.~\onlinecite{valley-Hall-BLMoS2,Morpurgo-3R-bilayer} 
it should be much smaller than the intrinsic SOC. 
We also neglect the difference between the effective masses of the two spin-split CBs at $U_g=0$ 
because it is quite small in MoS$_2$ 
and use a single effective mass $m_{\rm eff}$ for all bands. 
On the other hand, 
as shown in Fig.~\ref{fig:Berry-curv-calc}(b), for $U_g \lesssim 20$meV  
the $\mathbf{q}$ dependence of  $\Omega_{z,cb}(\mathbf{q})$
is  more important for 2H bilayers than  for 3R bilayers and we take it into account when we evaluate 
Eqs.~(\ref{eq:sigmavH}) and (\ref{eq:sigmasH}).

{We first assume that  $U_g\lessgtr \Delta_{cb}$ and  that 
 $E_F\sim$ few tens of \,meV. The case  $U_g=\Delta_{cb}$, which requires slightly 
different considerations, will be discussed below (see also Appendix \ref{sec:sigmaH-Appx}). 
Under the above assumptions and after summing up the contributions of all four bands shown 
in Fig.~\ref{fig:schem-Ug-2H}, one finds that} 
\begin{equation}
 \sigma_{v,H}^{2H}\approx\tau \frac{e^2}{\hbar}
 \left[\frac{\varepsilon_{cc}}{2\pi}\frac{U_g}{U_g^2-\Delta_{cb}^2}
 -\rho_{2d} U_g \left(\frac{\gamma_3}{\delta E_{bg}}\right)^2 \lambda_4(U_g)
 \right],
 \label{eq:sigmavH-2H}
\end{equation}
and 
\begin{equation}
 \sigma_{s,H}^{2H}\approx-\left[
 \frac{\varepsilon_{cc}}{2\pi}\frac{\Delta_{cb}}{U_g^2-\Delta_{cb}^2}
 +\rho_{2d} \Delta_{cb} \left(\frac{\gamma_3}{\delta E_{bg}}\right)^2 \lambda_4(U_g).
 \right]
 \label{eq:sigmasH-2H}
\end{equation}
Here $\varepsilon_{cc}=\frac{2 m_{\rm eff}}{\hbar^2} \gamma_{cc}^2$,  
$\rho_{2d}=m_{\rm eff}/2\pi\hbar^2$ is the two-dimensional density of states per spin and valley, and 
$\lambda_4(U_g)=\left(1+\frac{3}{4}\frac{\Delta_{vb}^2+t_{\perp}^2+U_g^2}{\delta E_{bg}^2}\right)$. 
One can see that $\sigma_{v,H}^{2H}$ vanishes for $U_g\rightarrow 0$, but $\sigma_{s,H}^{2H}$ remains finite.
When $U_g$ is of the order of $\Delta_{cb}$, the first term on the r.h.s of Eqs.~(\ref{eq:sigmavH-2H}) and 
(\ref{eq:sigmasH-2H}), which is related to the inter-layer contribution to the Berry curvature,  is larger than 
the second term.
{Moreover, this term changes sign as $U_g$ is changed from $U_g<\Delta_{cb}$ to $U_{g}>\Delta_{cb}$
and we expect that this leads to a sign change  in $\sigma_{v,H}^{2H}$ and  $\sigma_{s,H}^{2H}$. It is interesting to note 
that in e.g., lattice Chern insulators such a sign change of the off-diagonal conductivity was associated with a 
topological transition.  In our case  the sign change of  $\sigma_{v,H}^{2H}$ and  $\sigma_{s,H}^{2H}$ 
happens as the $\downarrow$ ($\uparrow$) bands first become degenerate 
at the $K$ ($-K$) point and then the degeneracy is lifted again as the electric field is increased further. }

Regarding  the $U_g=\Delta_{cb}$ case when two spin-polarized bands become degenerate  [see Fig.\ref{fig:schem-Ug-2H}(c)], 
in good approximation only the bands that remain non-degenerate have finite valley and spin Hall conductivity 
(see Appendix \ref{sec:sigmaH-Appx}).
Therefore the magnitude of  $\sigma_{v,H}^{2H}$ and 
$\sigma_{s,H}^{2H}$ is the same (apart from the fact that they are measured in different units).  
Summing up the contributions of the two $\uparrow$ ($\downarrow$)  bands in the $K$ ($-K$) valleys, 
one finds that  $\sigma_{v,H}^{2H}\approx \tau \frac{e^2}{\hbar} \tilde{\sigma}_{v,H}^{}$ and 
$\sigma_{s,H}^{2H}\approx \tilde{\sigma}_{v,H}^{}$, where    
\begin{equation}
 \tilde{\sigma}_{v,H}\approx\frac{1}{2}\left(\frac{\varepsilon_{cc}}{4\pi\Delta_{cb}}-\rho_{2d}\Delta_{cb}
 \left(\frac{\gamma_3}{\delta E_{bg}}\right)^2\lambda_{5}\right)
 \label{eq:tildesigmavH}
\end{equation}
and $\lambda_5=\left(1+\frac{3}{4}\frac{(\Delta_{vb})^2+t_{\perp}^2}{\delta E_{bg}^2}\right)$. 

{We do not discuss here the case when not all four spin-split bands  below the $E_F$ 
are occupied because we expect that  relatively large doping levels may be needed to suppress many-body effects, 
which are beyond the scope of the present work.}


\section{The $Q$ valleys}
\label{sec:Q-pt-main}

As one can see in, e.g.,  Figs.~\ref{fig:cryst-struct-bilayer}(a) and \ref{fig:cryst-struct-bilayer}(b), 
the local minimum of the CB at the $Q$ point of the BZ is almost degenerate with the $K$ valley, especially for 2H stacking. 
In our DFT band structure calculations the band edge is at the $K$ point for  both stackings 
and the $Q$ valleys would only be populated for a relatively strong $n$-doping.
We show the calculated $\delta E_{QK}$ values, i.e., the energy difference between the 
bottom of the $Q$ and the $K$ valleys, without/with taking into account SOC,  in  Table~\ref{tbl:dEKQ} below. 
\begin{table}[hbt]
\begin{tabular}{|c|c|c|}\hline
  & noSOC & SOC \\ \hline
2H & $9$ meV & $10$ meV\\  \hline
3R &  $75$ meV & $51$ meV\\ \hline
\end{tabular}
\caption{The calculated $\delta E_{QK}$ values for 2H and 3R bilayers 
without (no SOC) and with (SOC) taking into account the SOC.}
\label{tbl:dEKQ}
\end{table}
We note that in the case of monolayers it was found that $\delta E_{QK}$ depends
quite sensitively on the lattice constant, exchange-correlation potential\cite{2DMaterials-paper} 
and it may also depend on the level of theory (DFT or GW) used in the calculations. 
The same is expected to be  the case for  bilayers as well, where in addition the inter-layer separation 
used in the calculations may  also influence the location of the band edge.

Irrespective of the exact value of $\delta E_{QK}$ in DFT calculations,
it is of interest to understand if the 
six $Q$ valleys can affect the  valley Hall  conductivity  described in Section \ref{sec:valley-Hall} because 
strain or interaction with a substrate  may also affect energy difference between the bottom 
of the $K$ and $Q$ valleys. 
The calculations of Ref.~\onlinecite{JinboYang} indicate  that the Berry curvature is very small 
at the $Q$ point of monolayer TMDCs, therefore in our case it is only the inter-layer  contribution that 
needs to be considered. 
We find that, generally,  the Berry curvature  should be significantly smaller  in the $Q$ valley 
than in the $K$ valley  for bilayer MoS$_2$ (see Appendix \ref{sec:Q-pt-Appx}). 
{This is 
mainly because the bands are split by a momentum independent tunnelling  amplitude $t_{\perp,Q}$ which is 
much larger than the energy scale for momentum dependent coupling and the intra-layer spin-splitting.} 
Therefore, as long as inter-valley scattering between the $K$ and $Q$ valleys is not strong, 
the $Q$ valleys should have only a minor effect on the valley Hall and spin Hall  conductivities. 
{
Moreover, since the intra-layer spin-orbit coupling $\Delta_{Q}$ is one order of magnitude larger 
than $\Delta_{cb}$ at the $K$ point, we do not expect that in double gated devices a topological transition 
similar to the one at the $K$ point can take place.
}


\section{Discussion and summary}
\label{sec:conclusion}

In a very recent work\cite{KTLaw-spinHall} a different type of electrically controllable valley Hall effect,  
due to the Rashba type SOC, was  proposed in gated \emph{monolayer} TMDCs. 
In order to  obtain appreciable Rashba SOC in monolayer MoS$_2$, one would  need 
rather strong displacement fields\cite{PRX-paper} of the order of $0.3-0.4$\,eV\AA\cite{Qun-Fang},  
which are attainable, e.g.,  in ionic liquid gated devices. 
In contrast, given an inter-layer distance  of $d=2.975$\AA,\, a displacement field of $0.04$\,eV\AA\, would 
lead to an inter-layer potential difference $U_g\approx 13\,$meV which would give a roughly 
two-fold increase of $\Omega_{z,cb}$ in 2H bilayers with respect to the monolayer value. 
Thus we think that the Berry curvature is more easily tunable in bilayer TMDCs than in monolayers. 

Another way of investigating the Berry curvature may be offered by optical methods, 
where one can make use of the selection rules for circularly polarized light 
for intra-layer excitonic transitions at the $\pm K$ point. 
As it was shown in Refs.~\onlinecite{Srivastava,JZhou} for monolayer TMDCs, the Berry curvature 
acts as a momentum-space magnetic field and therefore it can split the energies of excitons 
that have non-zero angular momentum number. By extending this argument
to bilayers, one may expect that the Berry curvature should lead to a splitting of intra-layer 
excited excitonic states with non-zero angular momentum number and the effect would 
be  more pronounced, especially in 2H bilayers, than in monolayers. 
Note, that the Berry curvature  of both  the CB and the VB would contribute to this effect\cite{Srivastava,JZhou}. 
This would constitute a novel mechanism to influence intra-layer excitonic properties: the other layer does 
not only provide screening, but acts through the changing of the Berry curvature of the electrons 
and holes.

In summary, we have studied the Berry curvature properties and the corresponding valley Hall conductivities
of bilayer MoS$_2$. We have considered both 3R and 2H stacked bilayers and found intra-layer 
as well as inter-layer contributions to the Berry curvature, a situation not discussed before 
for layered materials.  Due to the inter-layer contribution, the Berry curvature is gate tunable. 
Moreover, we found that in  3R stacked bilayers  the  Berry curvature is much larger for 
electronic states in one of the layers than in the other one, i.e., it is effectively localized 
to one of the layers. For 2H stacking, on the other hand, it is usually much larger in the 
CB than in the VB, but it has the same magnitude in both layers. 
We  studied the  consequences  of the Berry curvature for n-doped samples. 
Firstly, the valley Hall conductivity will be finite if inversion 
symmetry is broken. Secondly, if the intrinsic SOC of the constituent layers 
is taken into account, the spin Hall conductivity is finite. 
Due to the SOC, in 3R bilayers all bands are non-degenerate and  spin-polarized, 
while in 2H bilayers the spin-polarized bands of the monolayer constituents are energetically degenerate 
as long as inversion symmetry is not broken. In 2H bilayers the interplay of 
SOC and finite interlayer potential can lead to a topological transition for one pair 
of spin-polarized bands. This leads to a change in the sign of the inter-layer contribution to the valley and 
spin Hall effects,  while the intra-layer contribution does not change sign. 
Our work highlights the role  of the stacking,  intra- and interlayer couplings on certain 
topological properties and can be relevant to 
a wide range of van der Waals materials.

\section{Acknowledgement}

{A. K. and G. B. acknowledge funding from FLAG-ERA through project ``iSpinText'' and from  SFB767.}
V. Z. and V. F. acknowledge support from the European Graphene Flagship Project,
the N8 Polaris service, and the use of the ARCHER supercomputer (RAP Project e547).


\appendix

\section{Derivation of $\mathbf{k}\cdot\mathbf{p}$ Hamiltonians of  bilayer TMDCs} 
 \label{sec:kp-general}

We  remind that the  symmetry properties of a band $\eta$ at a given $\mathbf{k}$-space point 
can be deduced by considering the transformation properties of Bloch states of the form \cite{PRB-paper}
\begin{eqnarray}
&|&\Psi_{\eta}^{t(b)}(\mathbf{k},\mathbf{r})\rangle \equiv
|\Psi_{l,m}^{t(b)}(\mathbf{k},\mathbf{r})\rangle \nonumber\\
 &=& \frac{1}{\sqrt{N}}\sum_{\mathbf{R}_n}e^{i \mathbf{k}(\mathbf{R}_n+\mathbf{t}_{Mo}^{t (b)})} 
Y_{l}^{m}(\mathbf{r}-[\mathbf{R}_n+\mathbf{t}_{Mo}^{t (b)})],
\label{eq:Bloch-wf}
\end{eqnarray}
where $Y_{l}^{m}(\mathbf{r})$ are rotating orbitals  formed from the atomic orbitals that contribute 
with large weight to a given band $\eta$ at a given $\mathbf{k}$-space point, $\mathbf{R}_n$ are lattice vectors in 
the direct lattice and $\mathbf{t}_{Mo}^{t (b)}$ give the positions of the Mo atoms 
in the top ($t$) and bottom ($b$) layers in the 2D unit cell. 
In the case of  2H bilayers  in external electric field the label $t$ corresponds to the layer at $+U_g$
potential, while the label $b$ to the layer at $-U_g$ potential. For zero electric field 
the labels $t$ and $b$ are somewhat arbitrary, nevertheless, for convenience we will keep them 
in the discussion that follows. 
(More rigorously, if the geometric position of the three-fold rotation axis is fixed, 
then each band can be labelled by an irreducible representation of the pertinent group of the wave vector.
Our choice of the lattice vectors and the coordinate system is shown below in Figure \ref{fig:lattice-geom}.)
 The situation is different for 3R bilayers: as explained in the main text, here the two layers are not equivalent 
 and one can define unambiguously the layer indices  $t$ and $b$.
\begin{figure}[htb]  
 \includegraphics[scale=0.4]{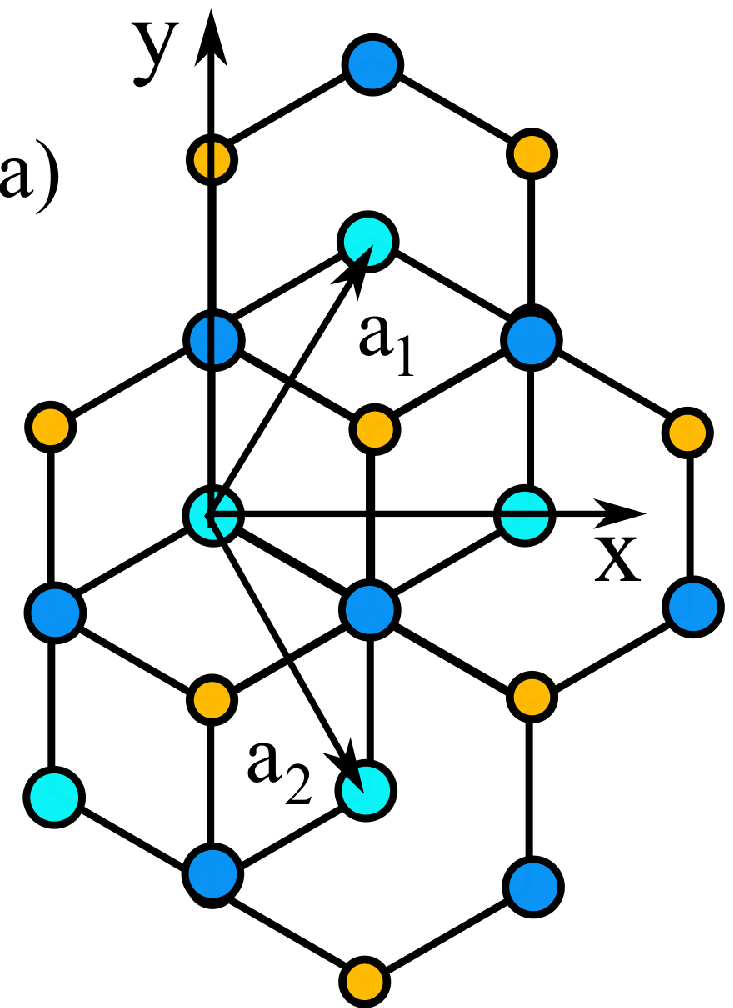}
 \hspace{1cm}
 \includegraphics[scale=0.4]{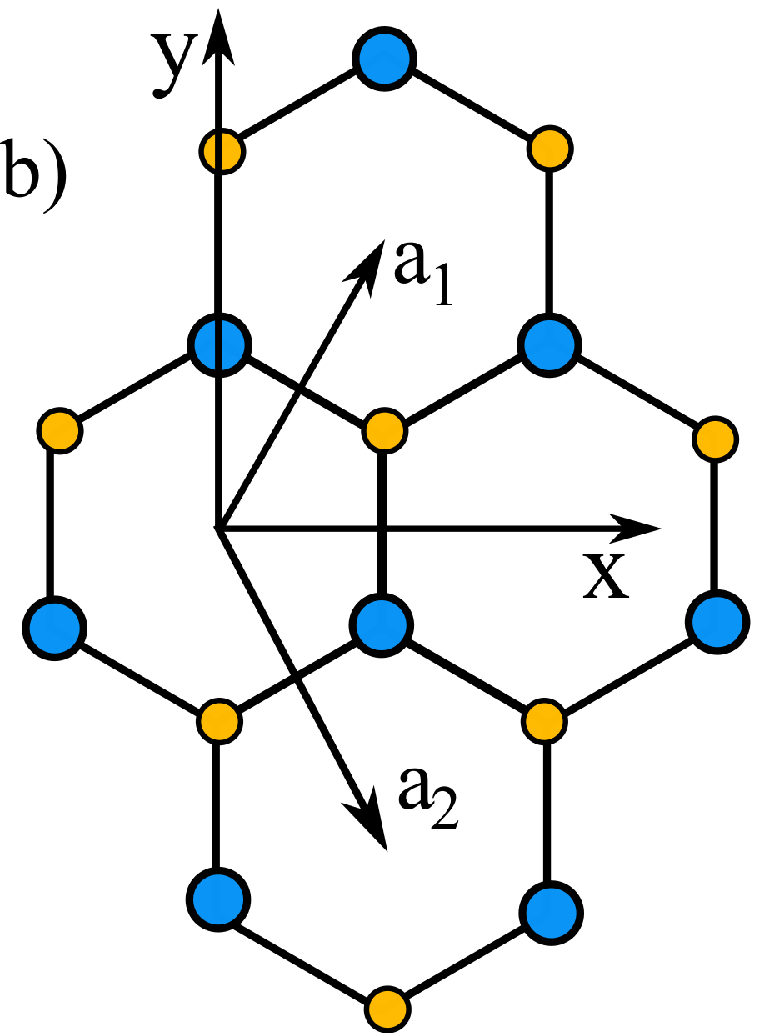}
 \caption{Schematic top view of the crystal lattice of a)  3R and b) 2H bilayer TMDC. 
   For 3R bilayers in a) the position of the metal atoms in the unit cell are  given by the vectors
   $\mathbf{t}_{Mo}^{t}=\frac{a}{2}(1,-1/\sqrt{3})^{T}$ and   $\mathbf{t}_{Mo}^{b}=(0,0)^{T}$. 
   Metal atoms in different layers are shown by different colours. 
   For 2H bilayers in b) only atoms in the top layer are visible. The position of the metal atoms 
   in the unit cell are given by the vectors 
   $\mathbf{t}_{Mo}^{t}=\frac{a}{2}(1,-1/\sqrt{3})^{T}$ and   $\mathbf{t}_{Mo}^{b}=\frac{a}{2}(1,1/\sqrt{3})^{T}$.
   Here $a$ is the lattice constant. 
 \label{fig:lattice-geom}}
\end{figure}

\begin{figure}[htb]
\includegraphics[scale=0.5]{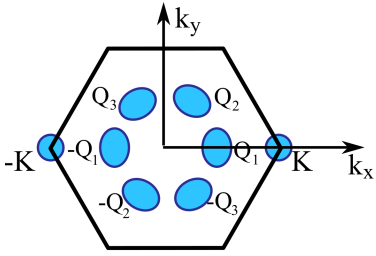}
\caption{The Brillouin zone with  the six $Q$ points and the $\pm K$ valleys.}
\label{fig:BZ-schem}
\end{figure}

The $\mathbf{k}\cdot\mathbf{p}$ Hamiltonian at a given $\mathbf{k}$-space point (see Fig.~\ref{fig:BZ-schem} for the 
Brillouin zone) can be then  
found by considering the transformation properties of the matrix elements 
$\langle \Psi_{\eta}^{\nu}(\mathbf{k},\mathbf{r})|\mathbf{\hat{p}}|\Psi_{\eta'}^{\nu'}(\mathbf{k},\mathbf{r})\rangle$,
where $\nu,\nu'=\{t,b\}$ and  $\mathbf{\hat{p}}=(\hat{p}_x,\hat{p}_y)$ are momentum operators 
(for details see, e.g., Refs.~\onlinecite{PRB-paper,2DMaterials-paper}). In this way one can arrive at a 
$7 \times 7$ 
model of a monolayer TMDCs\cite{2DMaterials-paper} and a corresponding 
$14 \times 14$ 
model for bilayers. 
Similarly, the SOC matrix elements can be found by considering 
$
\langle \Psi_{\eta}^{\nu}(\mathbf{k},\mathbf{r})|
\mathbf{\hat{L}}\cdot\mathbf{\hat{S}}
|\Psi_{\eta'}^{\nu'}(\mathbf{k},\mathbf{r})\rangle
$
where $\mathbf{\hat{L}}$ is a vector of angular momentum operators and $\mathbf{\hat{S}}$ is a vector of 
spin operators. 
In some cases, especially for effective Hamiltonians, 
it is easier to use the  theory of invariants\cite{BirPikus}. 
Both approaches  lead to the same results. 
We will use the following notation. The Pauli matrices $\sigma_{x,y,z}$ act in the space of top ($t$) and bottom ($b$) 
layer, while  the Pauli matrices $s_{x,y,z}$ act in the space of the spin degree of freedom.  
$\uparrow$ and $\downarrow$ denote  the eigenstates of $s_z$. Finally, the Pauli matrix 
$\tau_z$ describes the valley degree of freedom, and whenever  convenient, 
we use its eigenvalues $\tau=\pm 1$ for the same purpose.   


\section{ $\mathbf{k}\cdot\mathbf{p}$ Hamiltonians at the $\pm K$ point of the Brillouin zone}
\label{sec:K-pt}

\subsection{2H bilayer}
\label{subsec:K-pt-2H}

\subsubsection{$\mathbf{k}\cdot\mathbf{p}$ Hamiltonian}
\label{subsub:kp-K-pt-2H}

In the discussions below we will  refer to the group of the wavevector at the  
$\mathbf{k}=\pm K=\pm\frac{4\pi}{3 a}(1,0)^{T}$ 
points of the BZ,  which is ${D}_3$ for this stacking ($a$ is the length of the lattice vectors $\mathbf{a}_1$, and $\mathbf{a}_2$).
The character table of ${D}_{3}$ is given below in Table \ref{tbl:D3}.
\begin{table}[ht]
 \begin{tabular}{c|rrr}
   & E  & $2 C_3$  &  $3 C_2$\\ 
  \hline
  $\Gamma_1$ & 1  & 1 &  1\\
  $\Gamma_2$ & 1  & 1 & -1 \\ 
  $\Gamma_3$     & 2 & -1 & 0  \\
 \end{tabular}
 \caption{Character table of the point group ${D}_3$.} 
\label{tbl:D3}
\end{table}

We remind that  in monolayer TMDCs the  atomic $d_{z^2}$ orbitals of the metal atoms contribute with largest weight 
to the CB at the $\pm K$ points of the BZ.  Regarding the VB,  the $d_{xy}$ and $d_{x^2-y^2}$ orbitals are important 
at the $\pm K$ points. 
Taking into account that in 2H bilayers one of the monolayers is rotated by $180^{\circ}$ 
with respect to the other\cite{WangYao-bilayer}, one finds that the minimal basis set to describe the CB 
are the Bloch wavefunctions $|\Psi_{2,0}^{t(b)}(K,\mathbf{r})\rangle$ and for the VB the 
$|\Psi_{2,-2}^{t}(K,\mathbf{r})\rangle$,  $|\Psi_{2,2}^{b}(K,\mathbf{r})\rangle$
(This means that the top layer ``inherits'' the convention  we used in Refs.~\onlinecite{PRB-paper,2DMaterials-paper} 
for monolayer TMDCs, which is that at the $K$ point the Bloch wavefunction of the valence band is 
$|\Psi_{2,-2}^{}(K,\mathbf{r})\rangle$).

As a first step, let us neglect the SOC.
Using the coordinate system shown in Figure \ref{fig:lattice-geom}(b), 
one can easily show that $|\Psi_{2,0}^{t}(K,\mathbf{r})\rangle$ and $|\Psi_{2,0}^{b}(K,\mathbf{r})\rangle$
transform as partners of the two-dimensional irreducible representation (irrep) $\Gamma_3$ of ${D}_3$. 
This means that the CB is doubly degenerate at the $K$ point. We have 
checked that in our DFT calculations this is indeed the case (within numerical accuracy). 
Regarding the VBs, $|\Psi_{2,-2}^{t}(K,\mathbf{r})\rangle$ transforms as one of the partners 
of the irrep $\Gamma_3$, while $|\Psi_{2,2}^{b}(K,\mathbf{r})\rangle$ transforms as irrep $\Gamma_1$. 
Since the Bloch wavefunctions of the VBs of the $t$ and $b$ layers transform according to different 
irreps, the VB of the bilayer is not degenerate.   One can also notice that 
the operators $\hat{p}_{\pm}=\hat{p}_{x}\pm i \hat{p}_y$ also transform as partners of the irrep $\Gamma_3$.
Using the basis  $\{|\Psi_{\rm cb}^{t}\rangle, |\Psi_{\rm vb}^{t}\rangle, 
|\Psi_{\rm cb}^{b}\rangle,|\Psi_{\rm vb}^{b}\rangle \}$, 
the above considerations then lead to the following $\mathbf{k}\cdot\mathbf{p}$ Hamiltonian:
\begin{equation}
H^{2H}_{K} =
\left(
\begin{array}{cccc}
 \varepsilon_{cb}^{} & \gamma_3\,q_{+} &  \gamma_{cc}\,q_{-} &  \gamma_{cv}q_{+}\\
 \gamma_3\, q_{-} & \varepsilon_{vb}^{} &  \gamma_{vc}q_{+}  &  t_{\perp}\\
 \gamma_{cc}\,q_{+} &  \gamma_{vc} q_{-}  & \varepsilon_{cb}^{}  & \gamma_{3} q_{-} \\
 \gamma_{cv}q_{-}   &  t_{\perp}  &    \gamma_{3} q_{+}  &  \varepsilon_{vb}^{}      
\end{array}
\right).
\label{eq:H-2H-4dim-Appx}
\end{equation}
Here $t_{\perp}$ is a momentum independent tunneling amplitude between the VBs, $\gamma_3$ is the 
intra-layer coupling between the VB and the CB in each layer, while $\gamma_{cc}$ and $\gamma_{vc}=\gamma_{cv}$ 
are inter-layer couplings.
Here $q_{\pm}=\tau q_x \pm i q_y$ denotes the wavenumber measured from the $K$ (or $-K$)
point of the  BZ and $\tau=\pm 1$ is the valley index. 
A similar Hamiltonian to (\ref{eq:H-2H-4dim-Appx}), which only considered $t_{\perp}$ and 
neglected all other inter-layer coupling, was derived in Ref.~\cite{WangYao-bilayer}. 
We found  that  close to the $K$ point  the dispersion of the CB and VB  obtained from  DFT calculations 
can be fitted quite well by assuming that the inter-layer inter-band coupling 
constant $\gamma_{cv}$ is small and therefore we neglected this term. In contrast, the term $\sim \gamma_{cc}$ is 
needed
both to accurately fit the DFT band structure and for the Berry curvature calculations.


\subsubsection{Spin-orbit coupling}
\label{subsubsec:SOC-K-pt-2H}

For simplicity, we will only discuss here the case of zero external electric field.
Time reversal and inversion symmetries dictate  that  all bands are spin-degenerate throughout the BZ. 
In the simplest approximation we may take into account only the SOC in  the constituent monolayers. 
Using the basis 
$|\{\Psi_{\rm cb}^{t}\uparrow\rangle, |\Psi_{\rm cb}^{t}\downarrow\rangle, 
|\Psi_{cb}^{b}\uparrow\rangle,|\Psi_{cb}^{b}\downarrow\rangle \}$ for the CB (and an analogous basis set for the VB), 
the SOC Hamiltonian is 
\begin{equation}
 H_{\rm cb (vb),SOC}^{(1)}= \Delta_{cb (vb)}^{} \tau_z \sigma_z s_z 
 \label{eq:2H-SOC-1}
\end{equation}
Our  DFT calculations show  that the monolayer values $\Delta_{cb}$ and $\Delta_{vb}$ are indeed 
very close to the values  $\Delta_{cb}^{bl}$, $\Delta_{vb}^{bl}$ found in bilayers. 
The  term in Eqs.~(\ref{eq:2H-SOC-1}) is the most important one close to the $\pm K$ points. 

Strictly speaking, however,  $H_{\rm cb (vb),SOC}^{(1)}$ is not 
the only SOC term allowed by symmetries. 
Further terms can be obtained by using an  extended $\mathbf{k}\cdot\mathbf{p}$ model for the monolayers as in  
Ref.~\onlinecite{2DMaterials-paper}. For bilayers this extended basis contains $28$ basis states. 
The resulting SOC Hamiltonian has   matrix elements that connect  
basis states within the same layer as well as  inter-layer matrix elements. To simplify the discussion, 
we project the SOC onto the CBs and the VBs closest to the band gap. 
Then one can also use the theory of invariants to derive the SOC terms that appear in this effective low-energy model. 
For example, as already mentioned, the basis states   
$|\Psi_{2,0}^{t}(K,\mathbf{r})\rangle$ and $|\Psi_{2,0}^{b}(K,\mathbf{r})\rangle$
transform as partners of the two-dimensional irreducible representation $\Gamma_3$ of ${D}_3$. 
This means that for the CB  one can adapt the results derived 
for bilayer graphene\cite{Winkler-BLG,JFabian-bilayer,trilayer-SOC} and silicene\cite{Trauzettel-silicene}, 
where the low-energy sector of the Hamiltonian is also spanned 
by basis vectors transforming according to irrep $\Gamma_3$ of $D_3$. 
One finds that in lowest order of $\mathbf{k}$,  in addition to the  Eq.~(\ref{eq:2H-SOC-1}), 
one more term is allowed:
\begin{equation}
 H_{cb,SOC}^{(2)}=\Delta_{cb (vb)}^{(2)}\sigma_z(s_x q_y-s_y q_x). 
 \label{eq:2H-SOC-cb-2} 
\end{equation}
 Similarly to the CB, one finds that a term
\begin{equation}
 H_{vb,SOC}^{(2)}=\Delta_{vb}^{(2)}\sigma_z(s_x q_y-s_y q_x) 
 \label{eq:2H-SOC-vb-2} 
\end{equation}
can be added to the low-energy Hamiltonian in the VB. 
As one can see $H_{cb (vb),SOC}^{(2)}$  introduces a Rashba-like coupling within each of 
the layers.

We note that although  $H_{cb (vb),SOC}^{(2)}$ is  
diagonal in the layer space, a similar term is absent in monolayers. This follows from the different symmetries of 
monolayers and bilayers. In monolayer TMDCs the pertinent symmetry group   at $\pm K$ points is  the  $C_{3h}$ point group.  
This point group contains the symmetry element $\sigma_h$ corresponding 
to a horizontal mirror plane.  Polar vectors (such as $q_x$, $q_y$) and axial vectors (such as $s_x$, $s_y$) 
transform differently under $\sigma_h$ and therefore terms that contain their products, such as those in Eq.~(\ref{eq:2H-SOC-cb-2}), 
are not allowed. In the case of bilayer TMDCs the pertinent point group for 
the wavevector is $D_3$, which does not discriminate between polar and axial vectors and hence terms 
containing the products of wavenumber and spin components 
become admissible.

From a more microscopic point of view 
one can show that $H_{cb,SOC}^{(2)}$ and  $H_{vb,SOC}^{(2)}$ are both  due 
to an interplay of i) wavenumber dependent intra-layer coupling to higher or lower energy orbitals, and 
ii) certain off-diagonal intra-layer SOC matrix elements. The details of these calculations will be given elsewhere. 
The coupling constant $\Delta_{cb}^{(2)}$ and $\Delta_{vb}^{(2)}$ appear to be  small in MoS$_2$ 
and we could not reliably extract them from our DFT calculations.


\subsection{3R bilayer}
\label{subsec:K-pt-3R}

\subsubsection{$\mathbf{k}\cdot\mathbf{p}$ Hamiltonian}
\label{subsub:kp-K-pt-3R}

The 3R bilayer has lower symmetry than 2H bilayers,  e.g., as already mentioned in the main text, the crystal 
structure lacks inversion symmetry. The group of the wavevector at the  $\pm K$ points of 
the BZ  is $C_3$, for the character table see Table \ref{tbl:C3}.  
%
\begin{table}[htb]
 \begin{tabular}{c|rrr}
   & E & $C_3$ & $C_3^2$\\ 
  \hline
  $\Gamma_1$ & 1 & 1 & 1 \\
  $\Gamma_2$ & 1 & $\omega$ & $\omega^2$ \\ 
  $\Gamma_3$ & 1 & $\omega^2$ & $\omega$  \\
 \end{tabular}
 \caption{Character table of the group ${C}_3$. Here $\omega=e^{2 i \pi/3}$. 
\label{tbl:C3}}
\end{table}
One can see that all irreps of $C_3$ are one-dimensional. This  suggest that 
the bands of 3R bilayers are non-degenerate  in the $\pm K$ valleys.

Using the coordinate system shown in  Figure \ref{fig:lattice-geom}(a), 
the basis state  $|\Psi_{2,0}^{t}(K,\mathbf{r})\rangle$ 
($|\Psi_{2,0}^{b}(K,\mathbf{r})\rangle$) in the CB of the top (bottom) layer transforms as $\Gamma_2$  
($\Gamma_1$) of $C_3$. Regarding the VB,  
one finds that $|\Psi_{2,-2}^{t}(K,\mathbf{r})\rangle$ ($|\Psi_{2,-2}^{b}(K,\mathbf{r})\rangle$)  
transforms as $\Gamma_1$  ($\Gamma_3$). 
In the basis $\{|\Psi_{\rm cb}^{b}\rangle, |\Psi_{\rm vb}^{b}\rangle, 
|\Psi_{\rm cb}^{t}\rangle,|\Psi_{\rm vb}^{t}\rangle \}$ these symmetry considerations then lead to the 
following general form of the $\mathbf{k}\cdot\mathbf{p}$ Hamiltonian
\begin{equation}
H^{3R}_{K} =
\left(
\begin{array}{cccc}
 \varepsilon_{cb}^{b} & \gamma_3\,q_{+} &  \gamma_{cc}\,q_{-} &  t_{cv}\\
 \gamma_3\, q_{-} & \varepsilon_{vb}^{b} &  \gamma_{vc}q_{+}    &  \gamma_{vv}\,q_{-}\\
 \gamma_{cc}\,q_{+} &  \gamma_{vc}q_{-}  & \varepsilon_{cb}^{t}  & \gamma_{3} q_{+} \\
 t_{cv}   &  \gamma_{vv} q_{+}  &    \gamma_{3} q_{-}  &  \varepsilon_{vb}^{t}      
\end{array}
\right).
\label{eq:H-3R-4dim-Appx}
\end{equation}
Here we assumed that the diagonal elements  $\varepsilon^{t}$ and $\varepsilon^{b}$ 
can be different in the two layers. This can be motivated by noticing that 
the Mo  atoms in the two layers have different chemical  environment, since one of them is above 
a sulphur atom of the other layer, while the second Mo atom can be found in a hollow position. 
In contrast, in the crystal structure of 2H bilayers the metal atoms in  the two layers have the 
same chemical environment and therefore one  expects that the diagonal elements of the effective 
Hamiltonians are the same in the two layers, see Eq.~(\ref{eq:H-2H-4dim-Appx}). This argument can also be 
formulated from a symmetry point of 
view: in 2H bilayers the metal atoms are connected by symmetry operations of the crystal lattice, while 
this is not the case in 3R bilayers. 

Looking at Eq.~(\ref{eq:H-3R-4dim-Appx}), one can notice that the tunnelling amplitude $t_{cv}$, in principle, 
introduces a band repulsion between the CB of the bottom layer and the VB of the top layer even for $\mathbf{q}=0$. 
This looks  similar to the situation in 2H bilayers, where such a tunnelling element 
appears between the two VB states, see Eq.~(\ref{eq:H-2H-4dim-Appx}).  Indeed, in a recent work\cite{Xiaoou}
on the selection rules of optical transitions in 3R bilayers  an estimate of $t_{cv}\approx 50$ meV was given, 
which is comparable to $t_{\perp}$ in 2H bilayers. 
However, the analysis of our DFT calculations suggests  that $t_{cv}$ in 3R bilayers is much smaller than $t_{\perp}$.
To substantiate this claim  we show firstly the weight  of the atomic orbitals
in the highest energy VB of 2H bilayers at the $K$ point, as obtained from DFT calculations, 
in Table~\ref{tbl:2H-VB-atomweights}. 
Only  atomic orbitals  with non-zero weight are included.
\begin{table}[htb]
\begin{tabular}{c|ccccc}
            & $p_y$    & $p_x$     & $d_{xy}$    &  $d_{x^2-y^2}$ & tot\\ \hline
 Mo$^{(b)}$  &  $0.0$      &  $0.0$      &   $0.166$   &  $0.166$ &  $0.333$    \\
 Mo$^{(t)}$  &   $0.0$     &  $0.0$      &   $0.167$   &  $0.167$ &   $0.334$ \\
 S$^{(b)}_1$ &  $0.017$  &  $0.017$  &    $0.0$      &  $0.0$     &  $0.034$ \\
 S$^{(b)}_2$ &  $0.017$  &  $0.017$  &    $0.0$      &  $0.0$     &  $0.034$ \\
 S$^{(t)}_1$ &  $0.017$  &  $0.017$  &    $0.0$      &  $0.0$     &  $0.034$ \\
 S$^{(t)}_2$ &  $0.017$  &  $0.017$  &    $0.0$      &  $0.0$     &   $0.034$ \\
 \hline
\end{tabular}
\caption{The weight of the atomic orbitals in the highest valence band of 2H bilayer MoS$_2$ at the $K$ 
         point of the BZ. 
        Mo$^{(b)}$ (Mo$^{(t)}$) stands for the molybdenum atom in the top (bottom) layer. Similarly, 
         S$^{(b)}_1$ and S$^{(b)}_2$ (S$^{(t)}_1$ and S$^{(t)}_2$) stands for  the two 
         sulphur atoms in the two layers.}
\label{tbl:2H-VB-atomweights}         
\end{table}
As one can see, the atomic orbitals of both layers contribute with equal weight 
to this band. This agrees with the conclusion that one could draw from the Hamiltonian 
(\ref{eq:H-2H-4dim}): by diagonalizing it at $\mathbf{q}=0$, one can see that 
the VB states  of the two layers form ``bonding'' and ``antibonding'' 
states  due to the tunnelling $t_{\perp}$. In these new states the weight of the states from each 
layer is the same. 
Moreover, we find the same atomic weights as shown in Table~\ref{tbl:2H-VB-atomweights} for 
the second highest energy VB, which again supports the above interpretation. 

In the case of 3R bilayers, a similar argument  would suggest that 
both atomic orbitals belonging to the bottom layer and orbitals  belonging to the top layers would have 
finite weight in one of the CBs.  
In Tables~\ref{tbl:3R-CB-atomweights-1st} and  \ref{tbl:3R-CB-atomweights-2nd} we show the weight of the 
atomic orbitals in the first and second CB of 3R bilayer MoS$_2$, respectively. 
The SOC is neglected in these calculations since it is not important for the argument that we make.
\begin{table}[hbt]
\begin{tabular}{c|cccc}
             & $s$        &  $p_x$    &   $p_{y}$   &  $d_{z^2}$\\ \hline
 Mo$^{(b)}$  &  $0.0$   &  $0.0$      &   $0.0$     &  $0.0$     \\
 Mo$^{(t)}$  &   $0.043$      &  $0.0$      &    $0.0$    &     $0.743$  \\
 S$^{(b)}_1$ &  $0.0$     &  $0.0$  &    $0.0$  &  $0.0$   \\
 S$^{(b)}_2$ &  $0.0$     &  $0.0$  &    $0.0$  &  $0.0$   \\
 S$^{(t)}_1$ &  $0.0$     &  $0.017$    &    $0.017$      &  $0.0$   \\
 S$^{(t)}_2$ &  $0.0$     &  $0.017$    &    $0.017$      &  $0.0$   \\
 \hline
\end{tabular}
\caption{The weight of the atomic orbitals in the first 
         conduction band of 3R bilayer MoS$_2$ at the $K$ point of the BZ. 
         Mo$^{(b)}$ (Mo$^{(t)}$) stands for the molybdenum atom in the top (bottom) layer. 
         Similarly, S$^{(b)}_1$ and S$^{(b)}_2$ (S$^{(t)}_1$ and S$^{(t)}_2$) stands for  the two 
         sulphur atoms in the  top (bottom) layer.}
\label{tbl:3R-CB-atomweights-1st}
\end{table}
\begin{table}
\begin{tabular}{c|cccc}
             & $s$        &  $p_x$    &   $p_{y}$   &  $d_{z^2}$\\ \hline
 Mo$^{(b)}$  &  $0.043$   &  $0.0$      &   $0.0$     &  $0.744$     \\
 Mo$^{(t)}$  &   $0.0$      &  $0.0$      &    $0.0$    &     $0.0$  \\
 S$^{(b)}_1$ &  $0.0$     &  $0.016$  &    $0.016$  &  $0.0$   \\
 S$^{(b)}_2$ &  $0.0$     &  $0.017$  &    $0.017$  &  $0.0$   \\
 S$^{(t)}_1$ &  $0.0$     &  $0.0$    &    $0.0$      &  $0.0$   \\
 S$^{(t)}_2$ &  $0.0$     &  $0.0$    &    $0.0$      &  $0.0$   \\
 \hline
\end{tabular}
\caption{ The same as in Table \ref{tbl:3R-CB-atomweights-1st} but for the second 
          conduction band of 3R bilayer MoS$_2$ at the $K$ point of the BZ.
         }
\label{tbl:3R-CB-atomweights-2nd}
\end{table}
According to our  DFT calculations the atomic orbitals from the two layers are not admixed, i.e.,
the two layers are practically decoupled at the $K$ point. 
The  results in Table~\ref{tbl:3R-CB-atomweights-1st} and \ref{tbl:3R-CB-atomweights-2nd}  therefore suggest 
that $t_{cv}$, although allowed to be non-zero by symmetry considerations, is probably very small. 
We think  that the splitting of both the VB and CB states that can be clearly seen in Figure 1(a) of 
the main text is due to the difference between  the band edge energies $\varepsilon_{cb}^{b}$ and  $\varepsilon_{cb}^{t}$ 
($\varepsilon_{vb}^{b}$ and  $\varepsilon_{vb}^{t}$). 
This is the reason why we neglected $t_{cv}$ in the effective $\mathbf{k}\cdot\mathbf{p}$ model used in the main text.


\subsubsection{Spin-orbit coupling}
\label{subsubsec:SOC-K-pt-3R}

Since inversion symmetry is broken by the lattice in 3R bilayers, the bands need not be 
spin-degenerate when SOC is taken into account.
We will use the basis 
$\{|\{\Psi_{\rm cb}^{b}\uparrow\rangle, |\Psi_{\rm cb}^{b}\downarrow\rangle, 
|\Psi_{cb}^{t}\uparrow\rangle,|\Psi_{cb}^{t}\downarrow\rangle \}$ and an analogous basis for the VB.   
Considering only the SOC coupling of the constituent monolayers,  the SOC Hamiltonians are 
\begin{equation}
H_{cb (vb), SOC}^{(1)} = \Delta_{cb (vb)} \tau_z  s_z,
 \label{eq:3R-SOC-1}
\end{equation}
for the CB and the VB, respectively.
Note that,  in contrast to the 2H bilayers [see Eq.~(\ref{eq:2H-SOC-1})]  the Hamiltonian
(\ref{eq:3R-SOC-1}) does not depend on $\sigma_z$, i.e., it is  independent of the 
layer index. This is in agreement with the findings of Ref.~\onlinecite{Arita}.  
$H_{cb(vb),SOC}^{(1)}$  leads to the splitting of the otherwise spin degenerate
CB and VB  in each of the layers. 
These bands  are therefore non-degenerate  and  the spin-polarization of the bands is the same in both layers. 

Similarly to the 2H case, further SOC terms become possible if one considers virtual intra-layer and inter-layer processes.
To simplify the discussion, we project the SOC onto the CBs and the VBs closest to the band gap.  We list here the 
possible terms for the CBs, the same terms, albeit with different SOC strength, can be obtained for the VBs.  
Firstly, the intra-layer processes give rise to  a term similar to Eq.~(\ref{eq:2H-SOC-cb-2}):
\begin{equation}
H_{cb,so}^{(2,t (b))}=\Delta_{cb}^{(2,t(b))}(s_x q_y-s_y q_x), 
 \label{eq:3R-SOC-2}
\end{equation}
where in general the SOC coupling strengths are different in the two layers: $\Delta_{cb}^{(2,b)}\neq\Delta_{cb}^{(2,t)}$. 
Due to the lower symmetry of the 3R stacking, one finds further three non-zero inter-layer SOC terms. Defining 
$\sigma_{\pm}=(\sigma_x\pm i \sigma_y)/2$ and  $s_{\pm}^{\tau}=(s_x \pm i \tau s_y)/2$,  
one may write the first one as  
\begin{equation}
 H_{cb,so}^{(3)}=i\Delta_{cb}^{(3)}(\sigma_+ s_-^{\tau} - \sigma_- s_+^{\tau}) = 
 \frac{\Delta_{cb}^{(3)}}{2} (\tau_z \sigma_x s_y - \sigma_y s_x)
 \label{eq:3R-SOC-3}
\end{equation}
where $\Delta_{cb}^{(3)}$ describes direct spin-flip tunnelling between the CBs of the two layers. 
The second one reads 
\begin{eqnarray}
 &H_{cb,so}^{(4)}&=i\tau\Delta_{cb}^{(4)}(\sigma_+ s_+^{\tau} q_+^{\tau} - \sigma_- s_-^{\tau} q_-^{\tau}) \\ 
         &=& -\frac{\Delta_{cb}^{(4)}}{2} [ \sigma_x (s_x q_y + s_y q_x)+\tau_z \sigma_y (s_x q_x - s_y q_y)],\nonumber
 \label{eq:3R-SOC-4}
\end{eqnarray}
and the third one is 
\begin{equation}
  H_{cb,so}^{(5)}=\Delta_{cb}^{(5)}[\sigma_+ q_-^{\tau} + \sigma_- q_+^{\tau}] s_z =
  \Delta_{cb}^{(5)} [\sigma_x q_x + \tau_z \sigma_y q_y] s_z.
  \label{eq:3R-SOC-5}
\end{equation}
One can show that the last two terms, $H_{cb,so}^{(4)}$ and $H_{cb,so}^{(5)}$ are due to the interplay of a  
spin-dependent intra-layer hopping to a higher or lower energy orbital followed by a 
spin-independent inter-layer tunnelling or vice versa, a  spin-independent intra-layer hopping followed by a 
spin-dependent inter-layer tunnelling.

Our  DFT calculations suggest that in MoS$_2$  the terms corresponding to 
Eqs.~(\ref{eq:3R-SOC-2})-(\ref{eq:3R-SOC-5}) are much smaller than the monolayer SOC term 
Eq.~(\ref{eq:3R-SOC-1}). Moreover, we find that 
$\delta E_{cc}=(\varepsilon_{cb}^{b}-\varepsilon_{cb}^{t})/2\gg\Delta_{cb}^{(1)}$, 
which means that the low-energy CB bands are localized to the top layer. 
In contrast, $\delta E_{vv}=(\varepsilon_{vb}^{b}-\varepsilon_{vb}^{t})/2$ 
and $\Delta_{vb}^{(1)}$ are of similar magnitude in MoS$_2$  and following the convention 
of Ref.~\onlinecite{2DMaterials-paper}, whereby 
$\Delta_{vb}<0$ at the $K$ point, the highest energy state at $K$ is 
$|\Psi_{\rm vb}^{b}\downarrow\rangle$, followed by $|\Psi_{vb}^{t}\downarrow\rangle$, 
$|\Psi_{\rm vb}^{b}\uparrow\rangle,$ and finally $|\Psi_{vb}^{t}\uparrow\rangle$. 

Looking beyond MoS$_2$, in other MX$_2$ bilayers  it may happen that the crystal field splitting $\delta E_{cc}$ 
and the SOC strength $\Delta_{cb}^{(1)}$ are of comparable magnitude. In this case the two lowest energy CB bands would be  
localized on different layers, in the same way as in the VB of bilayer MoS$_2$.


\section{ $\mathbf{k}\cdot\mathbf{p}$ Hamiltonians at the $Q$ points of the Brillouin zone}
\label{sec:Q-pt-Appx}

In addition to the valleys at the $\pm K$ points, there are six $Q$ valleys in the CB, see 
Fig.~\ref{fig:BZ-schem}. 
Although in our DFT calculations the band edge in the CB can always be found  at the $\pm K$ point, 
the minima at the $Q$ points  is  close in energy to the minima at the $\pm K$ point 
[see Section \ref{sec:Q-pt-main} in themain text]. 
For finite doping therefore it may happen that both the $\pm K$ and the $Q$ valleys are populated. 
For this reason it is  of interest to understand the Berry curvature properties of the $Q$ valleys.  

Firstly, we note that the numerical calculations  of Ref.~\onlinecite{JinboYang} indicate that the Berry curvature 
in the CB of monolayer  MoS$_2$   is  much smaller at the $Q$ point than at the $\pm K$ point.
We will argue that this is also the case in bilayer TMDCs, 
i.e., in contrast to the $\pm K$ point, inter-layer coupling does not give a significant contribution 
to the Berry curvature in the Q valleys. 

To show this, we  remind that according to  DFT calculations (see, e.g., Ref.~\onlinecite{2DMaterials-paper}), 
in monolayer TMDCs the $d_{z^2}$, $d_{xy}$ and $d_{x^2-y^2}$ atomic orbitals of the metal atom have large 
weight at the $Q$ points. 
Therefore, in contrast to the $\pm K$ point where in the simplest approximation only the $d_{z^2}$ orbital
needs to be considered in the construction of Bloch wavefunctions, 
here it is necessary to take into account two other $d$ orbitals of the metal atom.
Considering, for concreteness, the Bloch wavefunction at the $Q_1$ point, it  can  be written:
\begin{eqnarray}
 |\Psi_{cb}(Q_1,\mathbf{r})\rangle = c_1 |d_{z^2}(Q_1,\mathbf{r})\rangle &+& 
 i c_2 |d_{xy}(Q_1,\mathbf{r})\rangle \\
 &+& c_3 |d_{x^2-y^2}(Q_1,\mathbf{r})\rangle \nonumber
\end{eqnarray}
where $|d_{z^2}(Q_1,\mathbf{r})\rangle$, $|d_{xy}(Q_1,\mathbf{r})\rangle$ and $|d_{x^2-y^2}(Q_1,\mathbf{r})\rangle$ 
are Bloch wavefunctions of the form shown in Eq.~(\ref{eq:Bloch-wf}) and $c_1$, $c_2$, $c_3$ are real numbers. 
The fact that $c_1$, $c_2$, $c_3$ are real can be shown  by noticing that the valleys  at $Q_1$ and $-Q_1$ are 
related by both time reversal $\mathcal{T}$ and the vertical reflection $\sigma_{v}^{y}$ with respect to the $y$ axis, 
see Figure \ref{fig:lattice-geom}(b). 
One can then use the  combined symmetry $\sigma_{v}^{y} \mathcal{T}$ to obtain restrictions on the wavefunction and 
hence on the coefficients  $c_1$, $c_2$, $c_3$. The same considerations apply to the $Q_2$ and $Q_3$ points as well.


\subsection{2H bilayer}
\label{subsec:Q-pt-2H}

Let us first assume zero external electric field and no coupling between the layers.
At the $Q_1$ point the small group of 
the wavevector is $C_2$, which contains the identity element and the  rotation $C_2^{x}$ by $\pi$  
around the $x$ axis, see Figure \ref{fig:lattice-geom}(b).  Bloch wavefunctions in the uncoupled 
top and bottom layers are also related by this symmetry and therefore they are given by 
\begin{widetext}
\begin{subequations}
\begin{align}
|\Psi_{cb}^{(t)}(Q_1,\mathbf{r})\rangle = c_1 |d_{z^2}^{(t)}(Q_1,\mathbf{r})\rangle + 
 i c_2 |d_{xy}^{(t)}(Q_1,\mathbf{r})\rangle + c_3 |d_{x^2-y^2}^{(t)}(Q_1,\mathbf{r})\rangle,
 \label{eq:Bloch-wf-Q-2H-t}\\
 |\Psi_{cb}^{(b)}(Q_1,\mathbf{r})\rangle = c_1 |d_{z^2}^{(b)}(Q_1,\mathbf{r})\rangle - 
 i c_2 |d_{xy}^{(b)}(Q_1,\mathbf{r})\rangle + c_3 |d_{x^2-y^2}^{(b)}(Q_1,\mathbf{r})\rangle.
  \label{eq:Bloch-wf-Q-2H-b}
 \end{align}
 \label{eq:Bloch-wf-Q-2H}
\end{subequations}
\end{widetext}
The minus sign appearing in the expression for  $|\Psi_{cb}^{(b)}(Q_1,\mathbf{r})\rangle$ with respect to 
$|\Psi_{cb}^{(t)}(Q_1,\mathbf{r})\rangle$ is due to the transformation rule $C_{2}^{x} d_{xy}= -d_{xy}$ of these 
atomic orbitals, while  $d_{z^2}$ and $d_{x^2-y^2}$ are not changed by $C_2^{x}$.

In the simplest approximation one may assume that to describe the low-energy states of  bilayer MoS$_2$ at the
$Q_1$ point it is sufficient to consider the states given by  Eqs.~(\ref{eq:Bloch-wf-Q-2H}). 
Neglecting, as a first step,  the SOC and up to second order in the wavenumber 
the corresponding $\mathbf{k}\cdot\mathbf{p}$ Hamiltonian $H_{Q_1}^{2H}$ for the bilayer case reads 
\begin{equation}
 H_{Q_1}^{2H}=\frac{\hbar^2 q_x^2}{2 m_{x,Q}}+ \frac{\hbar^2 q_y^2}{m_{y,Q}}+
 t_{\perp,Q}^{}\sigma_x + \gamma_x q_x \sigma_x+\gamma_y q_y \sigma_y.
 \label{eq:Q-noSOC-kp-2H}
\end{equation}
Here $q_{x,y}$ are measured from the $Q_1$ point and  
one can show that $t_{\perp,Q}^{}$ and $\gamma_{x,y}$ are real numbers. 
The first two terms  describe the dispersion at the $Q_1$ point of the isolated monolayers\cite{2DMaterials-paper}, 
$t_{\perp, Q}$ is a wavenumber independent inter-layer tunnelling, and the last two terms describe wavenumber dependent 
interlayer coupling.

Let us  consider the $\Gamma-K$ line of the BZ, where $q_y=0$. Neglecting, as a first step, the wavenumber dependent 
inter-layer coupling given by $\gamma_x q_x \sigma_x$, 
the spectrum of $H_{Q_1}^{2H}$ consists of two parabolas shifted in energy: 
$E_{\pm}= \frac{\hbar^2 q_x^2}{2 m_{x,Q}}\pm t_{\perp,Q}$. This allows to 
estimate the value of  $t_{\perp, Q}$  using the DFT band structure calculations and 
we find $t_{\perp, Q}\approx 205$\,meV.
If now  $\gamma_x q_x \sigma_x$ is taken into account,  by calculating  the eigenvalues of $H_{Q_1}^{2H}$ 
one can see that the minima of the two parabolas are not at $q_x=0$ 
but they can be found at slightly different ${q}_x$ points. This agrees with results of  the DFT 
band structure calculations. One can use this observation to extract the ratio $\gamma_x/t_{\perp, Q}\approx 0.82$\,\AA\, 
and one may use this value as an order of magnitude estimate for $\gamma_y/t_{\perp, Q}$ as well.

Regarding the SOC, we make the same approximation as for the $K$ point and  take into account only the 
intra-layer SOC of the constituent monolayers.  Thus  we use the Hamiltonian $\Delta_Q\tau_z \sigma_z s_z$, 
where  the SOC amplitude  $\Delta_Q\approx 70$\,meV is found from calculations in monolayer MoS$_2$\cite{2DMaterials-paper}. 
The SOC splits the CB in both layers but in 2H bilayers  the bands are spin-degenerate, as it is required 
by the time reversal and inversion symmetries. 
If  an interlayer potential difference $U_g$ is present due to an external electric field,
then the effective Hamiltonian reads  $\tilde{H}_{Q_1}^{2H}=H_{Q_1}^{2H}+ U_g \sigma_z + \Delta_Q \tau_z\sigma_z s_z$. 
Since inversion symmetry is broken, all bands are now spin $\uparrow$ or $\downarrow$ polarized. 
This behavior is qualitatively the same as for  the $K$ point, see Fig. 6(a) in the main text. 

Using  the eigenstates of $\tilde{H}_{Q_1}^{2H}$ to calculate the Berry curvature for the lowest-in-energy spin-split CB bands 
one finds  that 
$
|\Omega_{z,cb}^{(0)}(\mathbf{q}=0)| \sim \frac{\gamma_x \gamma_y}{t_{\perp, Q}^2} \frac{\Delta_Q\pm U_g}{t_{\perp,Q}}
$, 
where the $+$ ($-$) sign is for $\uparrow$ ($\downarrow$) polarized band (we remind that $\mathbf{q}$ is measured from 
the $Q_1$ point). 
Given the above estimate of $\frac{\gamma_x}{t_{\perp, Q}}$,  $\frac{\gamma_y}{t_{\perp, Q}}$ and $\Delta_Q/t_{\perp}$,   
$\Omega_{z,cb}^{(0)}$ is typically much  smaller than the corresponding inter-layer contribution 
in the $\pm K$ valley. 
On the other hand, as already mentioned, previous work\cite{JinboYang}   indicated  
that the intra-layer contribution is also very  small at the $Q$ point. We may thus conclude that even if 
both $Q$ and $K$ valleys are populated, the important contribution to the valley Hall effect should come
from the $K$ valleys.


\subsection{3R bilayer}
\label{subsec:Q-pt-3R}

The derivation of the effective Hamiltonian for the $Q$ valley in 3R bilayers is very similar to the 
case of 2H bilayers, see Section \ref{subsec:Q-pt-2H}. 
The general form of the Bloch wavefunctions at the $Q_1$ point of the isolated monolayers is 
\begin{widetext}
\begin{subequations}
\begin{align}
|\Psi_{cb}^{(t)}(Q_1,\mathbf{r})\rangle = c_1^{(t)} |d_{z^2}^{(t)}(Q_1,\mathbf{r})\rangle + 
 i c_2^{(t)} |d_{xy}^{(t)}(Q_1,\mathbf{r})\rangle + c_3^{(t)} |d_{x^2-y^2}^{(t)}(Q_1,\mathbf{r})\rangle
 \label{eq:Bloch-wf-Q-3R-t}\\
 |\Psi_{cb}^{(b)}(Q_1,\mathbf{r})\rangle = c_1^{(b)} |d_{z^2}^{(b)}(Q_1,\mathbf{r})\rangle + 
 i c_2^{(b)} |d_{xy}^{(b)}(Q_1,\mathbf{r})\rangle + c_3^{(b)} |d_{x^2-y^2}^{(b)}(Q_1,\mathbf{r})\rangle
  \label{eq:Bloch-wf-Q-3R-b}
 \end{align}
 \label{eq:Bloch-wf-Q-3R}
\end{subequations}
\end{widetext}
where  $c_1^{(t)}$, $c_2^{(t)}$ and $c_3^{(t)}$ in the top layer need not be exactly the same 
as $c_1^{(b)}$, $c_2^{(b)}$ and $c_3^{(b)}$ in the bottom layer. 
The effective Hamiltonian reads 
\begin{eqnarray}
 H_{Q_1}^{3R}=\frac{\hbar^2 q_x^2}{2 m_{x,Q}}+ \frac{\hbar^2 q_y^2}{m_{y,Q}} &+& \delta E_{cc,Q}\sigma_z\\
 &+&t_{\perp,Q}^{}\sigma_x + \gamma_x q_x \sigma_x+\gamma_y q_y \sigma_y,\nonumber
 \label{eq:Q-noSOC-kp-3R}
\end{eqnarray}
where $\delta E_{cc,Q}$ is the band-edge energy difference. This term is allowed since the top and bottom
layers are not related by any symmetry of the crystal lattice.

One can  again  consider the $\Gamma-K$ line of the BZ, where $q_y=0$.
By comparing the eigenvalues of $H_{Q_1}^{3R}$ with DFT band structure calculations, one can find that 
$\sqrt{t_{\perp,Q}^2+\delta E_{cc,Q}^2}\approx 170$\,meV.
The values of $\delta E_{cc,Q}$ and $t_{\perp,Q}$ cannot be extracted independently, but   
assuming a similar value for   $\delta E_{cc,Q}$ as at the $K$ point, where it is $\approx 30$\,meV, 
we obtain an estimate of $t_{\perp} \approx 167$\,meV.   Using  the 
difference between the positions of the band-edge minima along the $\Gamma-K$ line  one finds 
$\gamma_x/t_{\perp,Q}\approx 1.5$\,\AA\,\,  and 
one can take this value as an  estimate for  $\gamma_y/t_{\perp,Q}$ as well.

One can use the eigenstates of  $H_{Q_1}^{3R}$ to calculate the Berry curvature due to the inter-layer coupling. 
(In the approximation where only intra-layer SOC is taken into account, the energy scale $\Delta_{Q}^{}$
drops out from the calculations).
One finds that the inter-layer Berry curvature close to the Q valley minima is 
$|\Omega_{z,cb}^{0}(\mathbf{q}=0)| \approx \frac{\gamma_x \gamma_y}{t_{\perp, Q}^2} \frac{\delta E_{cc,Q}}{t_{\perp,Q}}
\approx 0.4$\,\AA$^2$,  
which is again significantly  smaller than the Berry curvature at the $\pm K$ points. 

We also note that according to our DFT calculations the energy difference $\delta E_{QK}$ is larger  in 3R bilayers  
than in 2H bilayers, see Section \ref{sec:Q-pt-main}. 
Therefore  the $Q$ valleys
would be populated only for stronger  doping. We may conclude that the contribution of the $Q$ valleys to 
the valley Hall conductivities should  be small in 3R bilayers.


\subsection{Remark on the model in Appendices \ref{subsec:Q-pt-2H} and \ref{subsec:Q-pt-3R}}
\label{subsec:Q-pt-3R-remark}

Looking at  Eqs.~(\ref{eq:Q-noSOC-kp-2H}) and (\ref{eq:Q-noSOC-kp-3R}), one would expect that inter-layer 
coupling would only weakly  affect the effective mass $m_{Q,x}$ (along the $\Gamma-K$ line) 
and therefore the two lowest energy bands at the $Q$ point would have equal effective masses.
To check the  validity of the two-band model introduced in Sections \ref{subsec:Q-pt-2H} and \ref{subsec:Q-pt-3R}, 
we have fitted the results of DFT band structure calculations to extract 
$m_{x,Q}^{(1)}$  and $m_{x,Q}^{(2)}$ for these two bands. For 2H bilayer 
the difference between $m_{x,Q}^{(1)}$ and $m_{x,Q}^{(2)}$ is around $4\%-5\%$,  while  it is  
$22\%-25\%$ for 3R bilayers. Within the $\mathbf{k}\cdot\mathbf{p}$ formalism such an effective mass difference can 
be understood as being due to coupling to other bands, not included into the simple two-band model.   
This indicates  the limitations of the two-band model. 


\section{Calculation of $\sigma_{v,H}^{2H}$ and $\sigma_{s,H}^{2H}$}
\label{sec:sigmaH-Appx}

We start by showing explicitly the results of the Berry curvature calculations. 
For concreteness, we first consider the $K$ valley. 
As explained in the main text, for $U_g>0 $ the four low energy CBs of the $K$ valley can be 
labelled by the spin index $\uparrow$, $\downarrow$ and by the index $\pm$  depending on whether the band edge 
can be found  at $\pm U_g$ potential at the $K$ point.  For spin $\uparrow$ bands one finds
\begin{subequations}
\begin{equation} 
 \Omega_{z}^{(0,\pm,\uparrow)}(\mathbf{q},U_g)=\mp \frac{1}{2}\left(\frac{\gamma_{cc}}{\Delta_{cb}+U_g}\right)^2 
 f_1^{3/2}(\mathbf{q},U_g), 
 \label{eq:Omega0up}
\end{equation}
\begin{equation} 
 \Omega_{z}^{(1,1,\pm,\uparrow)}(\mathbf{q},U_g)=\pm \frac{1}{2}\left(\frac{\gamma_{3}}{\delta E_{bg}}\right)^2 
 \lambda_{3}^{\uparrow}(U_g) f_1^{1/2}(\mathbf{q},U_g). 
 \label{eq:Omega11up}
 \end{equation}
\end{subequations}
The function $f_1(U_g)$ is defined as 
$
f_1(\mathbf{q},U_g)=\frac{1}{1+\left(\frac{\gamma_{cc} |\mathbf{q}|}{\Delta_{cb}+U_g}\right)^2}
$
and $\lambda_{3}^{(\uparrow)}(U_g)=1+\frac{3}{4}\frac{(\Delta_{vb}-U_g)^2+t_{\perp}^2}{\delta E_{bg}^2}$. 
For the spin $\downarrow$ bands  we assume that $U_g\neq \Delta_{cb}$ 
(the case  $U_g=\Delta_{cb}$  will be considered separately, see below), then   
the result is 
\begin{subequations}
\begin{eqnarray}
 \Omega_{z}^{(0,\pm,\downarrow)}(&\mathbf{q}&,U_g)=\mp {\rm sign}(U_g-\Delta_{cb})\times \nonumber\\ 
 &\frac{1}{2}&\left(\frac{\gamma_{cc}}{\Delta_{cb}-U_g}\right)^2 f_1^{3/2}(\mathbf{q},-U_g), 
\label{eq:Omega0down}
\end{eqnarray}
\begin{eqnarray}
 \Omega_{z}^{(1,1,\pm,\downarrow)}(&\mathbf{q}&,U_g)=\pm {\rm sign}(U_g-\Delta_{cb}) \times\nonumber\\
   &\frac{1}{2}&\left(\frac{\gamma_{3}}{\delta E_{bg}}\right)^2 \lambda_{3}^{(\downarrow)}(U_g) f_1^{1/2}(\mathbf{q},-U_g),
  \label{eq:Omega11down}
\end{eqnarray}
\end{subequations}
where ${\rm sign}[x]=1$ if $x>0$ and  ${\rm sign}[x]=-1$ if $x<0$ and 
$\lambda_{3}^{(\downarrow)}(U_g)=\lambda_3^{(\uparrow)}(-U_g)$. 

Repeating the calculations for  the $-K$ point, we find that the results can be written in the following form. 
Introducing the index $s=1$ ($s=-1$) for $\uparrow$ ($\downarrow$) spin-polarized  bands and $\tau=1$ ($\tau=-1$) for 
the $K$  ($-K$) valley,  for  $\tau\cdot s=1$ 
one finds 
\begin{subequations}
\begin{equation} 
 \Omega_{z}^{(0,\pm,s)}(\mathbf{q},U_g)=\mp  \frac{\tau}{2}\left(\frac{\gamma_{cc}}{\Delta_{cb}+U_g}\right)^2 
 f_1^{3/2}(\mathbf{q},U_g), 
 \label{eq:Omega0st1}
\end{equation}
\begin{equation} 
 \Omega_{z}^{(1,1,\pm,s)}(\mathbf{q},U_g)=\pm \frac{\tau}{2}\left(\frac{\gamma_{3}}{\delta E_{bg}}\right)^2 
 \lambda_{3}^{(s)}(U_g) f_1^{1/2}(\mathbf{q},U_g), 
 \label{eq:Omega11st1}
 \end{equation}
\end{subequations}
while for  $\tau\cdot s=-1$, 
\begin{subequations}
\begin{eqnarray}
 \Omega_{z}^{(0,\pm,s)}(&\mathbf{q}&,U_g)=\mp {\rm sign}(U_g-\Delta_{cb}) \times\nonumber\\
 &\frac{\tau}{2}&\left(\frac{\gamma_{cc}}{\Delta_{cb}-U_g}\right)^2 f_1^{3/2}(\mathbf{q},-U_g), 
  \label{eq:Omega0st-1}
\end{eqnarray}
\begin{eqnarray}
\Omega_{z}^{(1,1,\pm,s)}(&\mathbf{q}&,U_g)=\pm {\rm sign}(U_g-\Delta_{cb}) \times \nonumber\\
 &\frac{\tau}{2}&\left(\frac{\gamma_{3}}{\delta E_{bg}}\right)^2 \lambda_{3}^{(s)}(U_g) f_1^{1/2}(\mathbf{q},-U_g).
  \label{eq:Omega11st-1}
\end{eqnarray}
\end{subequations}

Turning now to the calculation of the valley Hall conductivities, for concreteness we again take the 
$K$ ($\tau=1$) valley. 
When $E_F>2(\Delta_{cb}+U_g)$ and therefore all four low-energy CB bands are populated, one may write 
\begin{widetext}
\begin{eqnarray}
 \tilde{\sigma}_{v,H}^{(0)}&=&\tilde{\sigma}_{v,H}^{(0,\uparrow)}+\tilde{\sigma}_{v,H}^{(0,\downarrow)}
= \sum_{n=\pm}\int\frac{d\mathbf{q}}{(2\pi)^2}
 \left[f_{n}^{(\uparrow)}(\mathbf{q})\Omega_{z}^{(0,n,\uparrow)}(\mathbf{q})
 +f_{n}^{(\downarrow)}(\mathbf{q})\Omega_{z}^{(0,n,\downarrow)}(\mathbf{q})\right],
 \label{eq:sigmavH0-integ} \\
 \tilde{\sigma}_{v,H}^{(1,1)}&=&\tilde{\sigma}_{v,H}^{(1,1,\uparrow)}+\tilde{\sigma}_{v,H}^{(1,1,\downarrow)}
=\sum_{n=\pm}\int\frac{d\mathbf{q}}{(2\pi)^2}
 \left[f_{n}^{(\uparrow)}(\mathbf{q})\Omega_{z}^{(1,1,n,\uparrow)}(\mathbf{q})
 +f_{n}^{(\downarrow)}(\mathbf{q})\Omega_{z}^{(1,1,n,\downarrow)}(\mathbf{q})\right], 
 \label{eq:sigmavH11-integ}
\end{eqnarray}
\end{widetext}
where $f_{n}^{(\uparrow,\downarrow)}(\mathbf{q})$ are Fermi-Dirac distribution functions. 
The valley Hall conductivity is then given by 
$\sigma_{v,H}^{2H}=\ \frac{e^2}{\hbar}(\tilde{\sigma}_{v,H}^{(0)}+\tilde{\sigma}_{v,H}^{(1,1)})$. 
Similarly, we may  define  
$\sigma_{s,H}^{(0)}=\tilde{\sigma}_{v,H}^{(0,\uparrow)}-\tilde{\sigma}_{v,H}^{(0,\downarrow)}$ and 
$\sigma_{s,H}^{(1,1)}=\tilde{\sigma}_{v,H}^{(1,1,\uparrow)}-\tilde{\sigma}_{v,H}^{(1,1,\downarrow)}$.  In terms of 
these quantities the spin Hall conductivity reads  $\sigma_{s,H}^{2H}=\sigma_{s,H}^{(0)}+\sigma_{s,H}^{(1,1)}$. 

We now explicitly calculate  $\tilde{\sigma}_{v,H}^{(0,\uparrow\downarrow)}$ and $\tilde{\sigma}_{v,H}^{(1,1,\uparrow\downarrow)}$. 
At zero temperature the integrals appearing in Eqs.~(\ref{eq:sigmavH0-integ}) and (\ref{eq:sigmavH11-integ}) are elementary.  
The upper limits of the integration, i.e., the Fermi momentum  $q_{F,\pm}^{(s)}$, 
$s=\uparrow,\downarrow$ can be found from 
the dispersion relations
\begin{subequations}
\begin{equation}
 E_F=\frac{\hbar^2 (q_{F,\pm}^{(\uparrow)})^2}{2 m_{\rm eff}}\pm \sqrt{(\Delta_{cb}+U_g)^2+(q_{F,\pm}^{(\uparrow)})^2 \gamma_{cc}^2}
\end{equation}
and 
\begin{equation}
 E_F=\frac{\hbar^2 (q_{F,\pm}^{(\downarrow)})^2}{2 m_{\rm eff}}\pm \sqrt{(\Delta_{cb}-U_g)^2+(q_{F,\pm}^{(\downarrow)})^2 \gamma_{cc}^2}.
\end{equation}
\end{subequations}

Using the notation $x=U_g-\Delta_{cb}$, one finds 
\begin{eqnarray}
 \sigma_{v,H}^{(0)}=\frac{1}{2} \frac{e^2}{\hbar} \frac{\varepsilon_{cc}}{2\pi} 
 \left[f_2(U_g)+ {\rm sign(x)} f_2(-U_g)\right],\\
 \sigma_{s,H}^{(0)}= \frac{\varepsilon_{cc}}{4\pi} 
 \left[f_2(U_g)- {\rm sign(x)} f_2(-U_g)\right],
 \label{eq:sigmavH0}
\end{eqnarray}
where $f_2(U_g)=\frac{1}{\sqrt{(\Delta_{cb}+U_g)^2+q_F^2\gamma_{cc}^2}}$, $q_F=\sqrt{\frac{2 m_{\rm eff}}{\hbar^2} E_F}$ 
and the energy  $\varepsilon_{cc}$ is defined as $\varepsilon_{cc}=\frac{2 m_{\rm eff}}{\hbar^2}\gamma_{cc}^2$. 
Using the value for $\gamma_{cc}=0.071$\,eV\AA\, that we obtained fitting our DFT band structure calculations and 
assuming, e.g.,  $E_F\sim 10 $\,meV,   one finds that typically  $q_F\gamma_{cc}\ll |\Delta_{cb}\pm U_g|$
and therefore   $q_F^{2}\gamma_{cc}^{2}$ can be neglected in $f_2(U_g)$ and $f_2(-U_g)$ (except when $U_g-\Delta_{cb}\approx 0$). 

Regarding $\sigma_{v,H}^{(1,1)}$ and $\sigma_{s,H}^{(1,1)}$, one obtains 
\begin{widetext}
\begin{eqnarray}
\sigma_{v,H}^{(1,1)}&=&-\frac{1}{2} \frac{e^2}{\hbar} \rho_{2d} \left(\frac{\gamma_3}{\delta E_{bg}}\right)^2
 \left[\lambda_3^{(\uparrow)}(U_g)[U_g+\Delta_{cb}]+\lambda_{3}^{(\downarrow)}(U_g)[U_g-\Delta_{cb}]\right]
 \label{eq:sigmavH11}
 \\
 \sigma_{s,H}^{(1,1)}&=&- \frac{1}{2} \rho_{2d} \left(\frac{\gamma_3}{\delta E_{bg}}\right)^2
 \left[\lambda_3^{(\uparrow)}(U_g) [U_g+\Delta_{cb}] - \lambda_{3}^{(\downarrow)}(U_g) [U_g-\Delta_{cb}] \right]. 
 \label{eq:sigmasH11}
\end{eqnarray}
\end{widetext}
Here $\rho_{2d}=m_{\rm eff}/2\pi\hbar^2$ is the two-dimensional density of states per spin and valley, 
$
\lambda_3^{(\uparrow)}(U_g) = 1+\frac{3}{4}\frac{(\Delta_{vb}-U_g)^2+t_{\perp}^2}{\delta E_{bg}^2}
$ 
and $\lambda_{3}^{(\downarrow)}=\lambda_{3}^{(\uparrow)}(-U_g)$.  Note, that 
$
\lambda_3^{(\uparrow)}+\lambda_{3}^{(\downarrow)}=2(\lambda_{3}(0)+\frac{3}{4}\frac{U_g^2}{\delta E_{bg}^2})=2 \lambda_4(U_g)
$ 
and 
$
\lambda_3^{(\uparrow)}-\lambda_{3}^{(\downarrow)}=-3 \frac{\Delta_{vb} U_g}{\delta E_{bg}^2}, 
$ 
hence for MoS$_2$  typically 
$
\lambda_3^{(\uparrow)}+\lambda_{3}^{(\downarrow)}\gg \lambda_3^{(\uparrow)}-\lambda_{3}^{(\downarrow)}
$
holds.   Therefore terms that are  $\sim \lambda_3^{(\uparrow)}-\lambda_{3}^{(\downarrow)} $ 
in Eqs.~(\ref{eq:sigmavH11}) and (\ref{eq:sigmasH11}) can be neglected.  
By repeating these calculations for the $-K$ ($\tau=-1$) valley, we arrive to the results given in the main text. 


Finally, we briefly discuss the $U_g=\Delta_{cb}$ case and for concreteness, we consider the $K$ valley. 
$\Omega_{z}^{(0,\pm,\uparrow)}$ and 
$\Omega_{z}^{(1,1,\pm,\uparrow)}$ [see Eqs.~(\ref{eq:Omega0up}) and (\ref{eq:Omega11up})] 
are smooth functions of $U_g$ and therefore one may write 
$
\tilde{\sigma}_{v,H}^{(0,\uparrow)}= \frac{1}{2} \frac{\varepsilon_{cc}}{2\pi} f_2(\Delta_{cb})
$
and 
$
\tilde{\sigma}_{v,H}^{(1,1,\uparrow)}=-\frac{1}{2}\Delta_{cb}\rho_{2d} \left(\frac{\gamma_3}{\delta E_{bg}}\right)^2 
\lambda_3^{(\uparrow)}(\Delta_{cb}).
$
Assuming  $E_F\sim 10 $\,meV,   one finds that typically  $q_F\gamma_{cc}\ll 2 \Delta_{cb}$ and therefore 
$f_2(\Delta_{cb})\approx \frac{1}{2\Delta_{cb}}$ and we may take 
$
\lambda_3^{(\uparrow)}(\Delta_{cb})\approx\lambda_3^{(\uparrow)}(0)=\lambda_5
$
because $\Delta_{cb}\ll\Delta_{vb}$ in MoS$_2$.  

Regarding the $\downarrow$ bands, since they are degenerate,  only $\Omega_{z}^{(1,1,\pm,\downarrow)}$
is finite. We repeat the calculations and find 
\begin{equation}
 \Omega_{z}^{(1,1,+,\downarrow)}= \Omega_{z}^{(1,1,-,\downarrow)}\approx \frac{1}{2} \left(\frac{\gamma_3}{\delta E_{bg}}\right)^2
 \frac{\Delta_{cb}+\Delta_{vb}}{2\delta E_{bg}}.
 \label{eq:Omega11down-deg}
\end{equation}
Strictly speaking, such term is also present for $U_g\neq \Delta_{cb}$, but it was  neglected in Eq.~(\ref{eq:Omega11down}) because 
it is much smaller than the one shown in Eq.~(\ref{eq:Omega11down}). 
We have also neglected  terms that are $\sim \mathbf{q}^2$. Using Eq.~(\ref{eq:Omega11down-deg})  to calculate 
$\tilde{\sigma}_{v,H}^{(1,1,\downarrow)}$ one obtains
\begin{equation}
 \tilde{\sigma}_{v,H}^{(1,1,\downarrow)}=\frac{1}{2} \left(\frac{\gamma_3}{\delta E_{bg}}\right)^2
 \frac{\Delta_{cb}+\Delta_{vb}}{2\delta E_{bg}}\left[\frac{(q_{F,+}^{(\downarrow)})^2+(q_{F,-}^{(\downarrow)})^2}{4 \pi}\right], 
\end{equation}
where  $q_{F,\pm}^{\downarrow}$ is determined by the dispersion relation 
$
E_F=\frac{\hbar^2 (q_{F,\pm}^{(\downarrow)})^2}{2 m_{\rm eff}}\pm\gamma_{cc} q_{F,\pm}^{(\downarrow)}. 
$
For $E_F\sim 10$\,meV one finds that 
$\tilde{\sigma}_{v,H}^{(1,1,\downarrow)}\ll \tilde{\sigma}_{v,H}^{(1,1,\uparrow)},
\tilde{\sigma}_{v,H}^{(0,\uparrow)}
$.


\bibliographystyle{prsty}

\begin{thebibliography}{10}
\bibitem{Viktor-DFT}
In our density functional calculations we used  the local density approximation. 
The structure of the monolayer was optimized as reported previously, see  \cite{2DMaterials-paper},  
while the inter-layer separation was taken from experiments and set to 6.147~\AA~ in both stackings. 
Since VASP relies on a plane-wave basis in a three-dimensional box, we set the vertical separation 
between the periodic images of the unit cell to 20 ~\AA, which is large enough 
to ensure that the structure can be considered isolated in two dimensions. 
We used a plane-wave cutoff energy of $600$eV and sampled the Brillouin zone with a $24 \times 24$ 
grid to compute the band structures.
\end{thebibliography}

\end{document}